\documentclass[aps,pra,twocolumn,showpacs,amsmath,amssymb,nofootinbib]{revtex4}

\usepackage[utf8]{inputenc}
\usepackage[english]{babel}
\usepackage[dvips]{graphicx}
\usepackage{amsfonts}
\usepackage{empheq}
\usepackage{graphicx}

\newcommand{\om} \omega   
\newcommand{\Om} \Omega
\newcommand{\eps} \epsilon

\newcommand{\be} {\begin{equation}}
\newcommand{\ee} {\end{equation}}
\newcommand{\ba} {\begin{eqnarray}}
\newcommand{\ea} {\end{eqnarray}}

\def\lrD{\mathrel{{\cal D}\kern-1.em\raise1.75ex\hbox{$\leftrightarrow$}}}
\def\lr #1{\mathrel{#1\kern-1.25em\raise1.75ex\hbox{$\leftrightarrow$}}}

\newcommand{\eqr}[1]{Eq.~(\ref{#1})}
\newcommand{\figr}[1]{Fig.~\ref{#1}}

\begin{document}

\title{Black-hole radiation in Bose-Einstein condensates}

\author{Jean Macher}
\email{jean.macher@th.u-psud.fr}
\affiliation{Laboratoire de Physique Th\'eorique, CNRS UMR 8627, B\^at. 210, Universit\'e Paris-Sud 11, 91405 Orsay Cedex, France}
\author{Renaud Parentani}
\email{renaud.parentani@th.u-psud.fr}
\affiliation{Laboratoire de Physique Th\'eorique, CNRS UMR 8627, B\^at. 210, Universit\'e Paris-Sud 11, 91405 Orsay Cedex, France}

\date{\today}

\begin{abstract}

We study the phonon fluxes emitted when the condensate velocity crosses the speed of sound, i.e., in backgrounds which are analogous to that of a black hole. We focus on elongated one dimensional condensates and on stationary flows. Our theoretical analysis and numerical results are based on the Bogoliubov-de~Gennes equation without further approximation. The spectral properties of the fluxes and of the long distance density-density correlations are obtained, both with and without an initial temperature. In realistic conditions, we show that the condensate temperature dominates the fluxes and thus hides the presence of the spontaneous emission (the Hawking effect). We also explain why the temperature amplifies the long distance correlations which are intrinsic to this effect. This confirms that the correlation pattern offers a neat signature of the Hawking effect. Optimal conditions for observing the pattern are discussed, as well as correlation patterns associated with scattering of classical waves. Flows associated with white holes are also considered.  
\end{abstract}

\pacs{03.75.Kk, 04.62.+v, 04.70.Dy}
\maketitle

\section{Introduction}

The analogy between sound propagation in 
nonhomogeneous media and light propagation in curved space-times 
has opened the possibility to detect the analogue of black hole radiation in the lab~\cite{Unruh:1980cg}. 
Indeed, when sound waves propagate in a flowing medium whose velocity crosses at some point the speed of sound, they experience the analogue of an event horizon. If the phonon state is stationary and regular across this sonic horizon,
one expects to obtain a thermal flux of phonons with a temperature $k_B T = \hbar \kappa/2\pi $, 
where $\kappa$ is the gradient of the flow velocity
evaluated at the sonic horizon. 
Since the analogy works perfectly in the hydrodynamical limit, 
the above result should be valid 
 at least when $\kappa$ is much smaller than the 
critical wave-vector characterizing the dispersion~\cite{Jacobson:1991bh,Unruh:1994je,Brout:1995wp,Corley:1996ar}.

Following the original work of Unruh, various setups were proposed, 
see \cite{Barcelo:2005fc} for a review. 
In Refs.~\cite{Garay:1999sk,Leonhardt:2003aa,Leonhardt:2004aa,Schutzhold:2006aa,Jain:2007ab,Wuester:2007aa,Balbinot:2007de,Carusotto:2008ep,Wuester:2008kh} 
the particular case of sound waves in dilute Bose-Einstein condensates (BEC) was considered.
These condensates have nice properties both from an experimental and a theoretical point
of view. From the first, progress in the manipulation and control of their physical properties is rapid, 
and from the second, the 
equations for the condensate and the phonons are well understood, as well as the 
approximations they involve~\cite{Dalfovo:1999zz}. 

In this work, we first aim to derive the properties of the phonon fluxes 
without making use of the gravitational analogy. 
In fact we also aim to determine the validity range of the analogy.
To achieve these goals, our analytical and numerical analysis are both based 
on the (exact) 
Bogoliubov-de Gennes (BdG) equation.
More specifically, we consider one dimensional stationary flows which contain one sonic horizon (black or white) surrounded by two infinite asymptotically homogeneous regions. In this case, at fixed conserved frequency $|\om|$, three types of asymptotic phonons exist, and a complete description of their scattering is given in terms of a $3 \times 3$ Bogoliubov transformation~\cite{Macher:2009tw}. 

Our second aim is to provide quantitative estimates of the spectral properties of the fluxes, in order to guide or explain experimental projects. To this end, we have performed a systematic numerical analysis.

Our third aim is to understand the links between local and nonlocal observables. In Gaussian states (e.g. vacuum and thermal states), the physical predictions encoded in a Bogoliubov transformation are all contained in two groups of expectation values: first occupation numbers (in the present case there are three of them, one for each type of outgoing phonons: $\bar n_\om^i = \langle a_{\om, \, i}^{out\dagger} \, a_{\om, \,  i}^{out} \rangle$, $i = 1,2, 3$),  and second, interference terms, such as $ \langle a_{\om, \, i}^{out} \, a_{\om, \,  j}^{out}  \rangle$ with $i \neq j$, which determine the long distance correlations~\cite{Parentani:2009va,Campo:2003gb,Campo:2003pa}. In this study, we were motivated by the recent work~\cite{Carusotto:2008ep} where a distinct pattern of density correlations was ``numerically observed'' when a sonic horizon forms. 
In what follows, we shall explain 
both the properties of this pattern, 
and also why, 
when taking into account the condensate temperature, 
the initial distributions of phonons tend to hide the 
black hole radiation in the occupation numbers whereas,  at the same time,
they reinforce the correlation pattern without affecting its spatial properties.
This confirms that the correlation pattern offers a neat signature of the Hawking 
effect~\cite{Balbinot:2007de}.

In Sec.~\ref{settings},  we derive the mode equation for the phonon field in one dimensional stationary condensates. We obtain an explicit 
fourth order equation which is valid for all nonhomogeneous condensates.
To be general, the sonic horizons we consider can be due either to the velocity flow $v(x)$, or to a varying sound speed $c(x)$, or to a combination of these two. We shall later see that some fluxes properties (such as the backscattering mixing left- and right-moving phonons) crucially depend on the particular combination which is realized. 

In Sec.~\ref{theory}, we analyze the mode equation, 
we identify the combinations of solutions which describe initial and final
asymptotic phonons, and relate them by the aforementioned 
$3 \times 3$ Bogoliubov transformation. In the next section, we show how it governs both local and nonlocal observables. This is carried out twice, without and with an initial temperature. In appendices, we consider the scattering of coherent states, which links the previous analysis with hydrodynamical experiments~\cite{Rousseau:2008aa}, and white holes. 

In Sec.~\ref{numres}, we numerically solve the mode equation and obtain the spectral properties
of the emitted  phonons. We consider both weakly and strongly dispersive regimes. The transition from one to the other
characterizes the validity range of the gravitational analogy.

\section{The settings\label{settings}}

\subsection{Dilute gases}

We give here the basic ingredients which describe the condensates and their linearized perturbations. 
The reader is referred to the review~\cite{Dalfovo:1999zz} 
for more details.

In a second quantized formalism, the set of atoms is described 
by a field operator $\Psi(t,{\bf x})$ which annihilates an atom at $t,{\bf x}$, and which 
obeys the equal time commutator
\be
[\Psi(t,{\bf x}), \Psi^\dagger(t,{\bf x}')] = \delta^3({\bf x}-{\bf x}')\, .
\ee
The time evolution of $\Psi$ is given by the Heisenberg equation
\be
i\hbar\,  \partial_t \Psi(t,{\bf x}) = [\Psi(t,{\bf x}),H],
\label{Heiseq}
\ee
where the Hamiltonian is
\be
H = \int d^3{x} \left\{   \frac{\hbar^2}{2m} \nabla_{\bf x} \Psi^\dagger \, \nabla_{\bf x}\Psi + V  \Psi^\dagger\Psi  + \frac{g}{2} 
 \Psi^\dagger\Psi^\dagger\Psi\Psi    \right\}.
\label{Hsc}
\ee
In this expression, $m$ is the mass of the atoms,  $V$ the external potential, 
and $g$ the effective coupling constant which describes the scattering of atoms by a local term.

At sufficiently low temperature, of the order of $300$~nK for $10^4$~atoms,
a large fraction of the atoms condense in a common state. %
To separate this state from its perturbations, the field operator 
is decomposed into a $c$-number wave describing the condensed atoms, $\Psi_0$, and a fluctuation: 
\be
 \Psi = \Psi_0 + \tilde \phi.
\label{c+f}
\ee 
In the mean field approximation, $\Psi_0$ satisfies the  Gross-Pitaevskii equation
\ba
i \hbar \partial_t \Psi_0 =  \left[  T  + V + g \rho_0 \right]\Psi_0,
\label{GP}
\ea
where the kinetic operator is $T = - \hbar^2 \nabla^2_{\bf x}/2 m$.
This equation 
guarantees that $\rho_0 = |\Psi_0|^2$ obeys the continuity equation:
$\partial_t \rho_0 + {\rm div} (\rho_0 {\bf v}) = 0$, where $\bf v$
is the condensate velocity.

\subsection{Stationary condensates}

In the general case, $V$ and $g$ depend on both $\bf x$ and $t$. In the body of this work we shall only consider stationary cases. In Appendix \ref{nonstatio} the time-dependent case is presented. Before proceeding, let us make clear that  stationarity means here that there is a Galilean frame (that needs not coincide with the lab frame) in which $V, g$ and therefore $\rho_0$  only depend on $\bf x$. From now on we work in this ``preferred'' frame. 

In this frame, the condensate wave function has the form 
\be
\Psi_0(t,{\bf x}) = e^{-i {\mu t/\hbar} } \times \sqrt{ \rho_0 ({\bf x})}\, e^{i W_0({\bf x})}, 
\label{eq1}
\ee
where $\rho_0 ({\bf x})$ gives the (mean) density of condensed atoms, 
$\mu$ is the chemical potential, and ${\bf k}_0(x)= \partial_{\bf x} W_0$ 
is the condensate wave vector. 

In the sequel, we shall work with one dimensional condensates. 
This means that the transverse excitations have energies  
much higher than the longitudinal ones, and than the interaction energy $g\rho_0$. 
The transverse excitations 
can therefore be ignored at sufficiently low energies. This simplifying 
hypothesis can be relaxed without significantly modifying
the forthcoming treatment.  
For one dimensional stationary condensates, 
the continuity equation reduces to
\be
 \rho_0 (x) v_{} (x) = {\rm cst.}
\label{cont}
\ee
where $v_{}(x) = \hbar k_0(x)/m$ is the velocity of the condensate, and
where $x$ is the longitudinal coordinate. 
Plugging \eqr{eq1} in \eqr{GP} gives
\be
\mu = \left[ \frac{m v^2(x)}{2}  
+ \rho_0^{-1/2}\,  T \rho_0^{1/2} 
+ V(x) + g(x)\rho_0(x) \right] . 
\label{tuned}
\ee

Because of \eqr{cont}, 
the nonhomogeneity of the background can be characterized by only two functions. We shall use $v_{}(x)$ and 
the $x$-dependent speed of sound 
\be
c^2(x) = g(x) \, \frac{\rho_0(x)}{m },
\label{cdef}
\ee
because the equation for the {\it relative} fluctuations only depend on these functions. 

\subsection{Bogoliubov-de~Gennes equation}

We show that the relative fluctuations obey, at the linear order, a fourth order equation which does not involve the external potential. 
Inserting \eqr{c+f} in \eqr{Heiseq}, and linearizing the equation in 
$\tilde \phi$, 
one gets 
\be
i\hbar \partial_t { \tilde \phi} = \left[  T  + V + 2 g 
|\Psi_0|^2
\right] { \tilde \phi} + 
g
\Psi_0^2
\,  { \tilde \phi}^\dagger . 
\label{BdG1}
\ee
Given the structure of this equation and \eqr{tuned}, 
we found that it is mathematically more convenient 
to work with the relative fluctuation $\phi$ defined by 
\be
 \Psi = \Psi_0 \, (1  +  \phi ).
\label{resc}
\ee
Then using Eqs.~(\ref{eq1},\ref{cdef}), one gets
\be
i\hbar \left(\partial_t + v \partial_x \right){ \phi} = 
T_\rho \, { \phi} + m  c^2
 \left[  {\phi} + {\phi}^\dagger \right],
\label{BdG2}
\ee
where we have introduced the ``dressed'' kinetic operator
\be
T_\rho = - \frac{ \hbar^2}{2 m \rho_0 }\, { \partial_x {\rho_{0}} }\, \partial_x 
=  - \hbar^2 \frac{ v_{}}{2 m}{ \partial_x \frac{1}{ v_{}} }\partial_x ,
\label{Trho}
\ee
which takes into account the nonhomogeneous character of the condensate density.
The last expression is obtained using \eqr{cont}, and is valid for stationary condensates only.

In this last case, the field operator can be written as
a superposition of the form 
\be\label{statio}
 \phi_\om(t, x) 
=  a_\om \,
 e^{- i \om t}\, \phi_\om(x)  + a_\om^\dagger \, \left(e^{-i \om t} \,\varphi_\om(x)\right)^*,  
\ee
where $ a_\om,   a_\om^\dagger$ are phonon annihilation and creation operators.
Inserting \eqr{statio} in  \eqr{BdG2}, and taking the commutator with both $a_\om$ and $a_\om^\dagger$ yields 
\be
\begin{split}
\left[ \hbar \left(\om + i  v_{} \, \partial_x \right) - T_\rho
- m  c^2  \right]\, \phi_\om =  m  c^2 \,  \varphi_\om, 
\\
\left[ -\hbar \left(\om + i v_{} \, \partial_x \right) - T_\rho 
- m  c^2  \right]\, \varphi_\om =  m  c^2 \,  \phi_\om .
\end{split}
\label{centraleq0}
\ee
It is instructive to obtain a single equation for $\phi_\om$ by eliminating $\varphi_\om$.
This can be done by dividing the first line by $c^2$ and acting on the resulting equation 
with the operator between brackets in the second line. After some manipulation, we obtain
\be
\begin{split}
\left\{
\left[ 
\hbar  \left(\om + i v_{} \, \partial_x \right) + T_\rho
\right]
\frac{1}{c^{2}} \left[ - \hbar \left(\om + i v_{} \, \partial_x \right) + T_\rho
\right] \right.\\
\left. -\hbar^2 { v_{}}{ \partial_x \frac{1}{v_{}} }\partial_x
\right\}
 \phi_\om = 0 .
\end{split}
\label{centraleq2}
\ee
This equation (or equivalently \eqr{centraleq0}) is valid for all stationary 1D condensates.
It contains no approximation besides those involved in the 
BdG equation~\cite{Dalfovo:1999zz,Wuester:2008kh}. 
In Appendix \ref{additional} we analyze its properties as well as its relations with other dispersive models.

\subsection{Near horizon trajectories and background profiles}

A sonic horizon can be obtained in two cases depending on whether it is $c + v_{}$ or $c - v_{}$ that crosses 
zero. Assuming it is $c + v_{}$, 
the condensate flows to the left, i.e., $v_{} < 0$. 
We further require that $c + v_{}$ smoothly crosses 
zero, i.e., the following expansion is valid in a finite range:
\be
c + v_{} = \kappa x + O(x^2).
\label{nhr}
\ee
We set to $x=0$ the location of the sonic horizon, and we call
the {\it near horizon region}, the range of $x$ where the neglect of nonlinear terms in 
\eqr{nhr} is valid.

To verify that this profile gives rise 
to a black hole horizon, we can either refer to the gravitational analogy~\cite{Barcelo:2005fc}, 
or directly analyze the characteristics of 
 Eq.~(\ref{centraleq2}). 
Adopting the second approach, we perform a long wavelength approximation to Eq.~(\ref{centraleq2}),
i.e., we drop the $T_\rho$ terms, and consider the WKB solutions 
of the resulting equation. 
In this approximation, right ($u$) and left ($v$) moving solutions (with respect to the condensate) 
decouple and are governed by $x$-dependent momenta $k^u_\om$, and $k^v_\om$ respectively. These are solutions to
\be
\om - k_\om\,  v_{}(x) = \pm k_\om \, c(x),
\label{uvdec2}
\ee
where the $+$ ($-$) sign governs $k_\om^u$ ($k_\om^v$), and where $c(x) > 0$. To get the space-time trajectories 
we consider the Hamilton-Jacobi equations ($dx/dt =  \partial \om/ \partial k , \,   
dk/dt = - \partial \om/ \partial x$).
These trajectories give the locus of constructive interference when considering
wave-packets.  In the near horizon region, for the right movers, \eqr{uvdec2} gives
\be
\begin{split}
&k= k_0 \, e^{- \kappa t},\\ 
&x= x_0 \,  e^{\kappa t}.
\end{split}
\label{nhr2}
\ee
By definition,
whenever such equations are obtained with $\kappa > 0$, 
one is dealing with a black hole horizon. 
When instead $\kappa < 0$,
the structure of the trajectories is that of a white hole. 
The two cases are related by a time reversal symmetry, see Appendix \ref{appWH}. 

In both cases 
the relevant quantity is the ``decay rate'' $\kappa$, 
given by the gradient
\be
\kappa = \left.\frac{d(c + v_{})}{dx}\right|_{\rm horizon} ,\label{kappadef}
\ee
evaluated at the sonic horizon $c +v_{} = 0$. 
(In the general relativistic jargon
it is called the ``surface gravity'' (when multiplied by $c$). 
It plays a crucial role in the laws of black hole thermodynamics~\cite{Bardeen:1973gs,Wald:1995yp},
and it fixes the temperature of black hole radiation when using relativistic fields~\cite{Hawking:1974sw}.) 
One should also point out that the left movers, the $v$-modes,
are hardly affected by the horizon since $ -c + v_{} \sim 2 v_{} - \kappa x$ 
is approximately constant in the near horizon region.

In the sequel, we shall work with 
\be
c(x) + v_{}(x) = c_0 \,
D \times {\rm sign}(x) \, \tanh^{\frac{1}{n}}\left[\left(\frac{|\kappa x|}{D c_0}\right)^n\right] .
\label{vdparam}
\ee
The parameter $1> D > 0$ determines the size of the 
 near horizon region, namely $|\Delta x |= D c_0/\kappa$. 
As we shall see, 
it plays a critical role in the deviations with respect to the standard
relativistic fluxes. 
The power $n$ controls the sharpness of the transition from the near horizon behavior 
to the asymptotic flat regions on either side. For $n \to \infty$, 
the transition becomes sharp. As discussed in~\cite{Corley:1996ar,Macher:2009tw},
sharp transitions give rise to nonadiabatic effects which produce 
superimposed oscillations on the fluxes. 
In this paper, we shall work with $n$ equal to 2,
and we refer to \cite{Macher:2009tw} for more details about this aspect.
\begin{figure}
\includegraphics{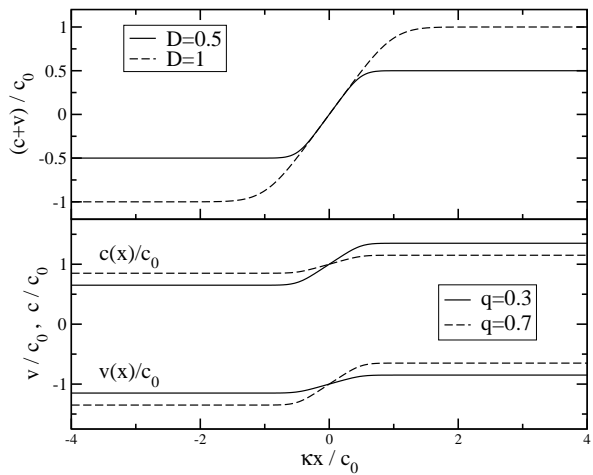}
\caption{Upper plot: shape of the profile $(c+v)/c_0$ as a function of $\kappa x/c_0$, for $D = 0.5$ (solid line) and $D=1$ (dashed line). Lower plot: separate profiles $c(x)/c_0$ and $v(x)/c_0$, for $D=0.5$ and $n=2$. The solid lines correspond to $q=0.3$ and the dashed lines to $q=0.7$. Both pairs of profiles give rise to the same function $c+v$, represented by the solid line in the upper plot.\label{fig::cd}}
\end{figure}

Given that the left-moving $v$-modes ``see'' the combination $c-v$,
and are coupled to the $u$-modes, we need to fix both $v$ and $c$,
and not only the combination of \eqr{vdparam}. To this end, we introduce a new parameter $q$ 
which specifies how $c+v$ is shared between $c$ and $v$: 
\be
\begin{split}
&c(x) = c_0 + (1- q) \, \left[c(x)+ v(x)\right] , \\  
&v(x) = - c_0 + q \,  \left[c(x)+ v(x)\right] .
\end{split}
\label{vdD}
\ee
For $q = 1$ (resp. $q = 0$) the hole is purely due to the gradient of $v$ (resp. $c$). 
In \cite{Carusotto:2008ep} the analysis was carried out with $q=0$, whereas in the recent experiment~\cite{Steinhauer:2009aa},
one finds $q \simeq 0.7$. In our numerical analysis we shall see that the value of $q$ 
has an important impact on the spectrum of the Hawking radiation. The influence of $D$ and $q$ on the functions $c(x)$ and $v(x)$ is illustrated in \figr{fig::cd}.

\section{Theoretical analysis\label{theory}}

To prepare the numerical analysis of \eqr{centraleq0}, it is worth studying the modes $(\phi_\om, \, \varphi_\om)$, and 
the Bogoliubov transformation relating the asymptotic in and out mode bases. 

We remind the reader that for a relativistic 2D massless field, the Bogoliubov transformation induced by propagating the field in \eqr{vdparam} is particularly simple because left and right moving modes decouple, see e.g.~\cite{Brout:1995rd}. When dealing with \eqr{centraleq2}, one faces two novelties. First, at fixed $\om$, four independent solutions now exist, and secondly, the left-right decoupling is no longer exact (even in the dispersionless limit $m \to \infty$).
Similar features have been already confronted in \cite{Corley:1996ar,Macher:2009tw}. 
However, the present case is more general, since both $v$ and $c$ vary. It is also more complicated, because $\phi$ and \eqr{centraleq2td} are complex,  
whereas the field and the mode equation in the former works were both real.

\subsection{Asymptotic solutions}

\begin{figure}
\includegraphics[scale=0.5]{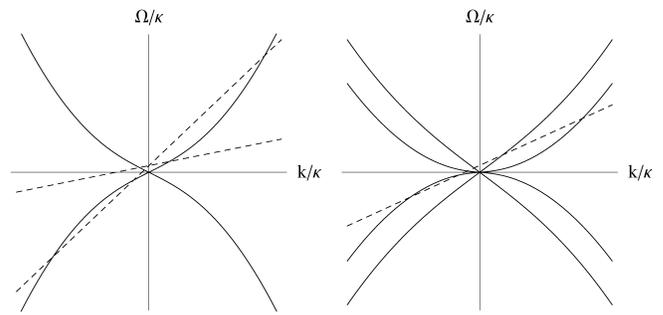}
\caption{\label{fig::reldispsols} Graphical resolution of the dispersion relation \eqr{reldisp}, for $q=1$ (left plot), and $q=0$ (rightplot). The straight lines represent $\om-v_\pm k$ and 
the curves represent $\Om_\pm(k)$. The solutions $k(\om)$ are given by the abscissa of their intersections.}
\end{figure}

In the above profiles, $c$ and $v$ are asymptotically constant in the regions $|\kappa x| \gg Dc_0$. 
In these regions, the general solution of \eqr{centraleq2} is a superposition of plane waves  $e^{i k x}$  with constant amplitudes. 
The asymptotic roots $k(\om)$ are solutions of (see \eqr{quarticdr})
\be
(\om - v_{\pm}  k )^2 
= c^2_{\pm} k^2  +  \frac{\hbar^2 k^{4}}{4 m^{2}} = \Omega^2_\pm(k).
\label{reldisp}
\ee
For $\om > 0 $, in the subsonic region $ c_+<\vert v_{+}\vert $, 
there is one real root $k^u_\om > 0$ which corresponds to a right mover
and another real root $k^v_\om <  0$ which describes a left mover, 
see \figr{fig::reldispsols}, where the two extreme cases $q=1$ (left plot) and $q=0$ (right plot) are represented.
There is also a pair of complex conjugated roots (since the equation is real). 
The root with a negative (resp. positive) imaginary 
part corresponds to a growing (resp. decaying) mode to the right.

\begin{figure}
\includegraphics[scale=0.8]{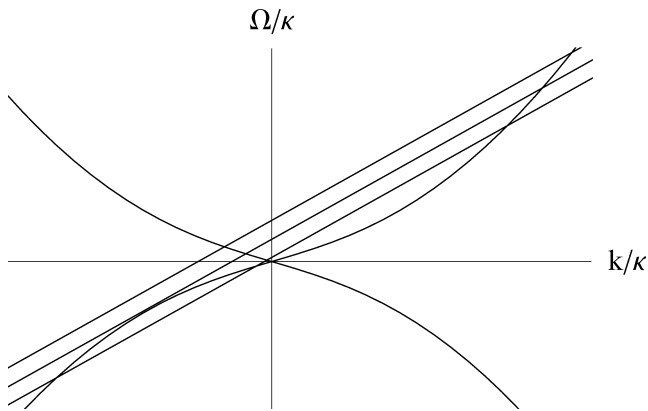}
\caption{Graphical resolution of the dispersion relation in the supersonic region 
for three values of $\om$: $0<\om <\om_{\rm max}$,
$ \om = \om_{\rm max}$
and $ \om>\om_{\rm max}$.\label{fig::merging}}
\end{figure}

In the supersonic region $ c_- > \vert v_{-} \vert$, 
for $\om$ larger than a critical frequency $\om_{\rm max}$ 
(we compute its value below)
there are only two real roots, 
as in a subsonic flow.
Instead for $0 < \om < \om_{\rm max}$ four real roots exist. This doubling of the number of real roots is illustrated in \figr{fig::merging}.
Thus the pair of growing and decaying modes which existed in 
the subsonic flow is replaced by a couple of oscillatory modes. Such a replacement 
is a generic feature of QFT in external fields, 
see \cite{Fulling:1989nb}.
In BEC it will occur at all 
horizons, for both black and white holes.

\subsection{Maximal frequency $\om_{\rm max}$}

The maximal frequency $\om_{\rm max}$ is the value of $\om$ where the two extra real roots merge.
It is thus reached when the straight line $\om - v_{-} k$ is tangent to $-\Om_-(k)$, the negative root of \eqr{reldisp} evaluated 
in the supersonic region. The corresponding value of $k_{\rm max}$ is 
\be
k_{\rm max} =\frac{1}{\sqrt 2} 
\frac{m}{\hbar }\sqrt{v_{-}^2 - 4c_-^2 + |v_{-}|\sqrt{v_{-}^2+8c_-^2}}.
\ee
Then $\om_{\rm max}$ is obtained by replacing $k$ by $k_{\rm max}$ in \eqr{reldisp}.
Using \eqr{vdparam} and \eqr{vdD}, it is thus of the form 
\be
\om_{\rm max} =    \Lambda \times  f(D, q),
\ee
where the ``healing'' frequency $\Lambda$ is related to the healing length computed with $c_0$ 
\be
\xi_0 =  \frac{\hbar}{ \sqrt{2}m c_0 },
\ee
by $\Lambda = \sqrt{2}c_0/\xi_0$. (The prefactor $\sqrt{2}$ has been added so that the quartic term in the dispersion relation~\eqr{reldisp} 
be equal to $k^4c_0^4/\Lambda^2$.) 
In what follows, $\Lambda$ (or the adimensional $\lambda=\Lambda/\kappa$) 
is referred to as the dispersion scale and is used to characterize the importance of dispersion.

\begin{figure}
\includegraphics[scale=0.8]{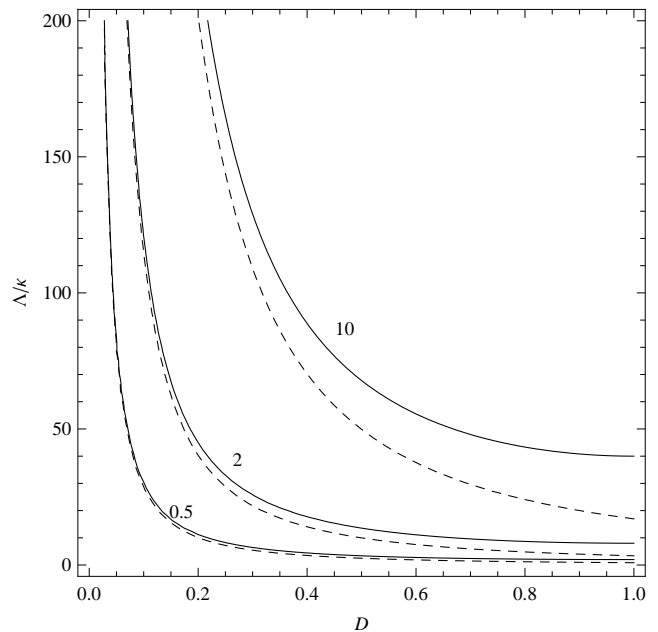}
\caption{Contours of constant $\om_{\rm max}/\kappa$ in the $(D,\lambda)$ plane, for $q=0$ (solid lines) and $q=1$ (dashed lines).\label{fig::ommaxcontours}}
\end{figure}

The contours of constant $\om_{\rm max}/\kappa$ in the $(D,\lambda)$-plane are shown in \figr{fig::ommaxcontours}, for $q=0$ and $q=1$. One sees that 
$q$ has only little influence on the value of $\om_{\max}/\kappa$ when this latter is small, but the effect becomes significant for larger values. The physical consequences of this shall be seen in 
Secs.~\ref{robustness}. 
Notice that for $D\ll 1$, one has $f(D,q)\propto D^{3/2}$. 
This means that $\om_{\rm max}$ can be much smaller than $\Lambda$.

\subsection[]{Mode orthonormality and mode completeness}

To proceed to the canonical quantization of $\phi$, one needs 
a mode basis which is orthonormal and complete.
The orthonormality is defined with respect to 
the conserved scalar product on the space of the solutions of \eqr{centraleq0}. 
In terms of the doublet $W_i = (\phi_i, \varphi_i)$ 
the scalar product takes the form~\cite{Leonhardt:2002aa}
\be
(W_1 \vert W_2) = \int_{-\infty}^{\infty}dx \, \rho_0(x) \left[ \phi_1^* \phi_2 - \varphi_1^* \varphi_2\right].\label{BECnorm}
\ee
The presence of $\rho_0$ in this product follows from the use of 
the rescaled fluctuations 
defined in \eqr{resc}. Similarly, the equal time commutator reads
\be
[ \phi(t,x), \phi^\dagger (t,x')] = \frac{1}{\rho_0(x)} \delta(x-x') .
\label{ETC} \ee
This $x$-dependent measure induces no difficulty, and moreover one can get rid of it by using the non-Cartesian coordinate $y$ defined by $dy = dx \rho_0(x)\label{ydef}$. 

When the condensate is homogeneous the quantization is rather straightforward
and well known~\cite{Dalfovo:1999zz}. However, when the flow $c+v$ possesses a horizon, the situation is more subtle.
Therefore, before considering the inhomogeneous backgrounds of \eqr{vdparam}, 
let us quantize $\phi$ when $c$ and $v_{}$ are constant both in space and time.
The complete description we shall obtain transposes to a BEC that of Ref.~\cite{Macher:2009tw}.

\subsubsection[]{Homogeneous condensates, $k$-representation}

In homogeneous condensates, it is appropriate to express the field in terms of exponentials $e^{ikx}$ and creation/destruction operators labeled by the (real) wave-vector $k$,
\be
\hat \phi = \int_{-\infty}^\infty {dk }
\left[  \hat a_k  \, \phi_k + 
\hat a_k^\dagger  \, \varphi_k^* \right],
\label{phiink}
\ee
where
\be
\phi_k(t,x) = \frac{ e^{- i \om_k t + i kx }}{\sqrt{2 \pi \rho_0}} \, {\mathfrak u}_k , \quad 
\varphi_k(t,x) = \frac{e^{- i \om_k t + i kx }}{\sqrt{2 \pi \rho_0}}  \, {\mathfrak v}_k ,
\ee
and where $\om_k$ is the positive solution of \eqr{reldisp}. 
Using these expressions, and introducing $\bar W_k = (\varphi_k^*, \phi_k^*)$,
the ``bar'' doublet associated with $W_k = (\phi_k, \varphi_k)$, one verifies that the orthonormality conditions 
\be
\begin{split}
&(W_k \vert W_{k'}) = -(\bar W_k \vert \bar  W_{k'})= \delta(k-k'),\\ 
&(\bar W_k \vert  W_{k'}) = 0,
\end{split}
\label{Dnorms}
\ee
are satisfied when the amplitudes $\mathfrak{u}_k, \mathfrak{v}_k$ take their standard value~\cite{Abrikosov:1976aa}, irrespectively of the (sub- or supersonic) value of the condensate velocity $v_{}$. Explicitly, using \eqr{centraleq0} and \eqr{quarticdr}, one obtains
\be
\begin{split}
&\mathfrak{v}_k = D_k \, \mathfrak{u}_k, \quad \,\vert \mathfrak{u}_k \vert^2 - \vert \mathfrak{v}_k \vert^2 = 1,
\\
&D_k  =  \frac{1}{mc^2}\left[ \hbar \sqrt{c^2\, k^2 + \frac{\hbar^2 k^4}{4m^2}} - \left(m  c^2 + \frac{\hbar^2 k^2}{2m}\right)   \right].
\end{split}
\label{D}
\ee
The fact that $D_k$ is independent of the condensate velocity $v$
follows from Galilean invariance, 
the condensate being homogeneous.
Indeed, in the coordinates $t_c, x_c$ comoving with the fluid, related to the coordinates $t,x$
by $t_c = t$, $x_c = x - vt$, the 2D wave vector of components $(\om_k,k)$ in the $t,x$ system 
has comoving components $(\Om_k = \om_k - vk, k)$ 
where all reference to $v$ drops out when using $k$ to label modes, 
as can be seen from \eqr{reldisp}. 

The operators $\hat a_k, \hat a_k^\dagger$ obey the usual bosonic commutation relation 
$[\hat a_k, \hat a_{k'}^\dagger] = \delta(k-k')$. 
The relationships between mode doublets $W_k$ and these operators 
follow from
\be
\begin{split}
\hat a_k  &= (W_k \vert \hat W),\\ 
\hat a_k^\dagger &= - (\bar W_k \vert \hat W),
\end{split}
\label{aoverl}
\ee
where the doublet operator is $\hat W = (\hat\phi,\hat \phi^\dagger)$.

In this $k$-representation, one can also easily verify that the mode basis is complete. That is, 
starting with the commutators $[\hat a_k, \hat a_{k'}^\dagger] = \delta(k-k')$
and the doublets $W_k$, making use of the completeness 
(in the sense of Fourier analysis) of the exponentials $e^{ikx}$
with $k$ real from $[-\infty, \infty]$, and using that 
$D_k = D_{-k}$ (which is a necessary condition)
one verifies that \eqr{ETC} is satisfied, again irrespectively of the value of the condensate velocity $v_{}$.

Therefore, when using $\om$ instead of $k$ to label modes and operators, one should discard the growing and the decaying solutions of \eqr{centraleq2}. 
Unlike the oscillatory plane waves they cannot be reached in a limiting procedure, starting with square integrable functions. 

\subsubsection[]{Nonhomogeneous condensates, $\om$-representation}

When considering nonhomogeneous but stationary flows, one should use the conserved frequency $\om$  to label modes. 
Then one notices that the change of variable $k\to \om$ in \eqr{phiink} proceeds very differently 
according to the sub- or supersonic character of the flow. As a preliminary step, we consider separately sub- and supersonic homogeneous flows. 
A superscript $u$ ($v$) shall be added to characterize right (left) movers with respect to the condensate. 

In a homogeneous subsonic flow, only two real roots of \eqr{reldisp} exist: $k^u(\om)>0$ and $k^v(\om)< 0$. Given that $dk /d\om$ 
never crosses zero, one can re-express \eqr{phiink} as 
\be
\phi(t,x) = \int_0^\infty d\om \left[ e^{- i \om t} \,  \hat \phi_\om(x) +  
e^{ + i \om t} \, \hat \varphi_\om^\dagger(x) 
\right] ,
\label{sumoverom}
\ee
where
\be
\begin{split}
&\hat \phi_\om(x) = \hat a^u_\om \, \phi^u_{\om}(x) + \hat a^v_\om \, \phi^v_{\om}(x), \\
&\hat \varphi_\om(x) = \hat a^{u}_\om \, \varphi^{u}_{\om}(x) + \hat a^{v}_\om \,   \varphi^{v}_{\om}(x).
\end{split}
\label{just2}
\ee
The rescaled modes and operators are $\phi_\om = \phi_k \sqrt{dk/d\om}$, and
$\hat a_\om = \hat a_k \sqrt{dk/d\om}$. One easily verifies that 
the factors 
$\sqrt{dk/d\om}$ 
guarantee that all 
$\delta(k-k')$ obtained in the former subsection
are consistently replaced by $\delta(\om-\om')$.  
That is, the 
 operators $\hat a_\om, \hat a^\dagger_\om$ obey 
$[\hat a_\om, \hat a_{\om'}^\dagger] = \delta(\om - \om')$.
Similarly the modes $\phi^u_{\om}$ and $\phi^v_{\om}$ 
are orthogonal to each other, 
and possess unit positive  norm (in the sense of a Dirac distribution $\delta(\om - \om')$).

We now consider a homogeneous supersonic flow. Starting again from \eqr{phiink} one decomposes, as in \eqr{sumoverom},  the field as an integral over $\om$ of a sum of right- and left-moving modes. When considering the left-moving sector in left-moving flows, $dk /d\om$ does not cross zero. Therefore, as in subsonic flows, all left-moving (positive norm) modes can still be monotonically labeled by $\om$ belonging to $[0, \infty]$. The same is no longer true for the right-moving sector. In fact, the integral $\int_0^\infty dk$ splits into an integral over $\om$ belonging to $[0, \infty]$ plus another piece over negative frequencies belonging to $[-\om_{\rm max}, 0]$. This last interval terminates at $-\om_{\rm max}$ where the two new real roots $k^u(\om) > 0$ merge. 

Thus, for $ \om > \om_{\rm max}$, $\hat \phi_\om, \hat \varphi_\om$ read as in \eqr{just2} since one has only one (positive norm) $u$-root. Instead, when $0 < \om < \om_{\rm max}$, three real $u$-roots exist: the continuation (in $\om$)  of this positive norm one, plus two new roots with negative $\Om$, see Fig.~\ref{fig::merging}. In this case, one has 
\begin{align}
\hat \phi_\om(x) &\equiv \int \frac{dt}{2 \pi } e^{i \om t} \, \phi(t,x)   \nonumber\\
 &= \hat a^u_\om \, \phi^u_{\om}(x) + \hat a^v_\om \, \phi^v_{\om}(x) + \sum_{l=1,2} \hat a^{u \, \dagger}_{-\om, l}\,  \left[\varphi^u_{-\om, l}(x)\right]^*, \label{modeinom2}\\
\hat \varphi^\dagger_\om(x) &\equiv \int \frac{dt}{2 \pi } e^{-i \om t}\,  \phi(t,x)  \nonumber\\
 &= \hat a^{u\, \dagger}_\om \left[ \varphi^{u}_{\om}(x)\right]^* + \hat a^{v\, \dagger}_\om \left[\varphi^{v}_{\om}(x)\right]^* + \sum_{l=1,2} \hat a^{u}_{-\om, l}\,  \phi^u_{-\om, l}(x).
\end{align}
In the first equation, a complex conjugate and a subscript $-\om$ have been used 
to characterize the two new modes. This means that the two doublets  $W^u_{-\om, l}= (\phi^u_{-\om, l}, \varphi^u_{-\om, l})$ 
have a positive norm and obey \eqr{centraleq0} with a frequency $i \partial_t = -\om < 0$. It should also be noticed that the above operators $\hat \phi_\om(x), \hat \varphi_\om(x)$ contain both annihilation and creation sectors. This allows one to write $\phi(t,x)$ as in \eqr{sumoverom}, i.e. as an integral over $\om\in \, [0, \infty]$. 

In metrics which contain a transition from a subsonic to a supersonic flow, because of the scattering on $v(x)$, the modes that are bounded on one side of the horizon are generally not bounded on the other side. However, given the linearity of \eqr{centraleq0}, one can always construct bounded modes as linear combinations of the above modes, requiring that the coefficient of the growing mode be zero~\cite{Macher:2009tw}. 

In fact, when $\om<\om_{\rm max}$, three independent bounded combinations can be constructed and $\hat \phi_\om$, instead of \eqr{modeinom2}, now reads: 
\be
\hat \phi_\om(x) = \hat a^u_\om \, \phi^u_{\om}(x) + 
\hat a^v_\om \, \phi^v_{\om}(x) + 
 \hat a^{u \, \dagger}_{-\om}\,  \left[\varphi^u_{-\om}(x)\right]^*,
\label{modeinom}
\ee
and similarly for $\hat \varphi_\om(x)$. In this expression, $\phi^u_{\om}$, $\phi^u_{-\om}$ and $\varphi^u_{-\om}$ stand for yet unspecified, normalized bounded modes. The particular cases of the in and out bases 
shall be defined in the next section. For $\om > \om_{\rm max}$ instead, there is one growing mode on each side of the horizon. Hence only 
{\it two} independent bounded linear combinations can be constructed, so that $\hat \phi_\om$ is still decomposed as in \eqr{just2}.

\subsection[]{Bogoliubov transformation}

\subsubsection{Vacuum instability and spontaneous pair production}

Since we are dealing with a stationary situation, the energy of the state of $\phi$ is constant. Hence, one might have thought that no phonons could possibly be spontaneously emitted. This is not the case, because the vacuum is unstable against the production of phonon pairs when two conditions are met. First, negative energy excitations must exist. We saw in the last section that it is the case when the flow is supersonic, for $\om < \om_{\rm max}$. 

However, this is not enough, as can be understood by considering a homogeneous condensate propagating in a frictionless translation faster than sound in the lab frame. One also needs spatial gradients to define unambiguously the unique ``preferred'' stationary frame, and to couple the so-defined negative energy excitations to positive energy ones. When these conditions are met, the Hamiltonian of $\phi$  has the following structure
\ba
H &=& \int_0^{\om_{\rm max}}  d\om \, \hbar \om \left( \hat a^{u\, \dagger}_\om \hat a^{u}_\om 
+ \hat a^{v\, \dagger}_\om \hat a^{v}_\om  
-   \hat a^{u\, \dagger}_{-\om} \hat a^{u}_{-\om}\,\right)
\nonumber \\
&& + \int_{\om_{\rm max}}^\infty  d\om \, \hbar \om \left( \hat a^{u\, \dagger}_\om \hat a^{u}_\om 
+ \hat a^{v\, \dagger}_\om \hat a^{v}_\om  \right).
\label{HK}
\ea
From this expression three conclusions can be drawn. First, phonons with $\om > \om_{\rm max}$ cannot participate in the pair production since no partner with the corresponding negative frequency exists. Second, both $u$ and $v$ positive frequency phonons participate in the vacuum instability. 
Which of the two channels contributes most depends on the intensity of the coupling with the negative frequency excitations. Third, the diagonalization of $H$ in the sector $0< \om <\om_{\rm max}$ is ambiguous because by a unitary ($\om$-diagonal) Bogoliubov transformation, the above quadratic form is left unchanged. 

Therefore, one needs an additional physical criterion to remove this ambiguity. In the general (time-dependent) case no such criterion exists, and the notion of {\it phonon} is inherently ambiguous~\cite{Birrell+Davies,Wald:1995yp}. However, in the present case of stationary asymptotically homogeneous condensates, there is no ambiguity to define two complete sets of modes. The first set of modes (the ``in'' modes)  characterizes the initial phonons propagating towards the sonic horizon. The second set, the ``out'' modes, characterizes the asymptotic particle content of the scattered field configurations. Having the two sets, we can compute how they are related and how this relationship governs the decay of the vacuum.

\subsubsection{Space-time structure of in and out  wave-packets}

The procedure to identify the in and out modes is 
standard~\cite{Brout:1995rd}. One should construct ``broad wave packets'',
i.e., superpositions of stationary modes, so as to extract the asymptotic temporal behavior by looking at 
the stationary phase condition $\partial_\om S = 0$. This equation is equivalent to 
Hamilton's equation since the mode phase $S$ coincides with the Hamilton-Jacobi
action in the WKB approximation. From this asymptotic behavior, one identifies the modes  (in fact the
doublets $W_{\om, \, a}= (\phi_{\om, \, a}, \varphi_{\om, \, a})$) 
that are associated with each initial (and final) one-phonon state. 
It should also be pointed out that these modes 
acquire a precise physical meaning when considering coherent states, see App.~\ref{coher}.

\begin{figure}
\includegraphics{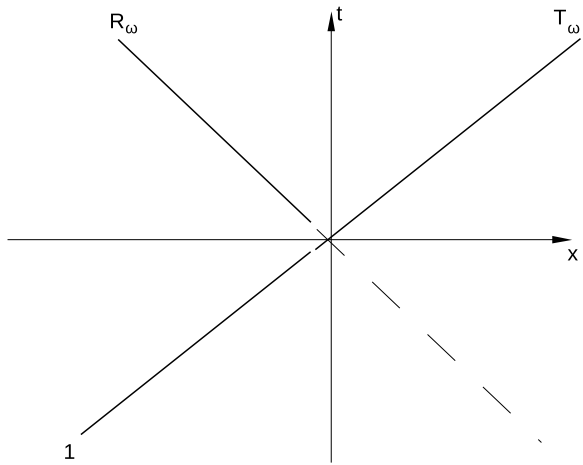}
\caption{Schematic space-time representation of a wave-packet made out of modes $\phi^{u,in}_\om$, for $\om>\om_{\rm max}$. At early times, the packet is purely right-moving. It is then scattered into a transmitted right-moving part and a reflected left-moving one. The dashed line represents the initially unexcited ``ancestor'' of the late-time left-moving packet.\label{fig::uinomRT}}
\end{figure}

\begin{figure}
\includegraphics{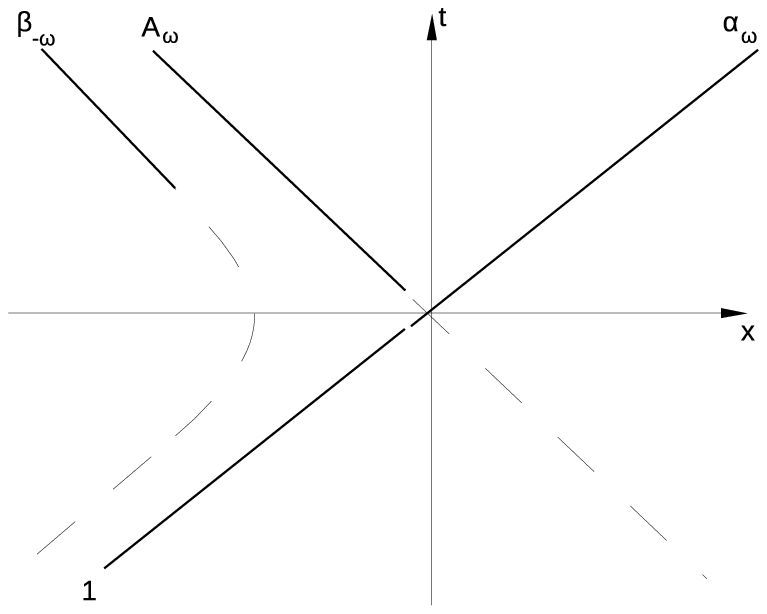}
\caption{Same as \figr{fig::uinomRT} for $\om<\om_{\rm max}$, when pair production occurs. \label{fig::uinom}}
\end{figure}

It is useful to visualize these modes. 
Let us first describe the in mode $\phi^{u, in }_{\om}$, i.e., the particular solution of \eqr{centraleq2}
which contains in the past only a right-moving packet. We shall do it twice, both for $\om> \om_{\rm max}$,
where there is only some elastic scattering (see \figr{fig::uinomRT}), and for $\om < \om_{\rm max}$, in the presence of pair creation (see \figr{fig::uinom}). 
In both cases, initially, one only has the incoming branch with unit norm. At late time, 
when $\om > \om_{\rm max}$, one has a
transmitted $u$-mode with amplitude $T_\om$,
and a reflected $v$-mode with amplitude $R_\om$. For $\om < \om_{\rm max}$, in addition to these modes
there is the negative frequency mode $(\varphi^{u, out }_{-\om})^*$. The description of the other in and out modes is obtained without difficulty.

\subsubsection[]{Bogoliubov transformation}

From \eqr{HK} one sees that for $\om > \om_{\rm max}$ 
one has only two modes with positive frequency/energy. 
Because the condensate is not translation
invariant, there is some scattering. Therefore, in and out
$\phi$ modes are related by 
\be
\begin{split}
\phi^{u, in }_{\om} &= T_{\om} \, \phi^{u, out}_{\om} 
+ R_{\om} \,  \phi^{v, out}_{\om},\\
\phi^{v, in }_{\om} &= -  R^*_{\om}\,  \phi^{u, out}_{\om} + T^*_{\om} \, \phi^{v, out}_{\om}.  
\end{split}
\label{RT}\ee
The other modes, the $\varphi_\om^{u/v, \, in/out}$, are related among themselves by the same relations.
Since there is no mixing of $\phi_\om$ with $\varphi_\om^*$,
the conservation of the norm trivially implies $|T_{\om}|^2 + |R_{\om}|^2= 1$. 
One is thus dealing with an elastic scattering between $u$ and $v$ modes,
and $T_{\om}$, $R_{\om}$ are the transmitted and reflected amplitudes, respectively.
This is a ``trivial'' transformation in the sense that there is
no spontaneous pair production. The vacuum of these high frequency modes 
is thus stable.

It should be pointed out that this elastic scattering 
is also found, for all modes, when the flow remains everywhere subsonic. 
In fact, because of dispersion, the high frequency modes
with  $\om > \om_{\rm max}$ do not see/experience the presence of the sonic horizon. 
This can be verified by computing the trajectories followed by 
wave-packets centered around some $\om > \om_{\rm max}$. 

The situation is radically different for $\om < \om_{\rm max}$. In this case the mode mixing
is nontrivial, and three equations are needed 
to characterize the transformation 
\ba
\phi^{u, in }_{\om} &=& \alpha_{\om} \, \phi^{u, out}_{\om} + \beta_{-\om}   \left(\varphi^{u, out}_{-\om}\right)^*
+ \tilde A_{\om} \,  \phi^{v, out}_{\om},
\nonumber \\
\phi^{v, in }_{\om} &=& \alpha^v_{\om} \, \phi^{v, out}_{\om} + 
B_\om   \left(\varphi^{u, out}_{-\om}\right)^*
+  A_{\om}\,  \phi^{u, out}_{\om},\label{Bog}\\
\phi^{u, in }_{-\om} &=& \alpha_{-\om} \, \phi^{u, out}_{-\om} + \beta_{\om} 
  \left(\varphi^{u, out}_{\om}\right)^*
+ \tilde B_{\om} \left( \varphi^{v, out}_{\om}\right)^* .\nonumber
\ea
The coefficients are given by the overlap of the corresponding (normalized) in and out doublets, 
e.g. 
\be
\beta_{-\om}= - (\bar W^{u, out}_{-\om}\vert W^{u, in }_{\om}).
\label{betao}
\ee 
The normalization of the coefficients 
then immediately follows, e.g. the first equation 
(together with the corresponding one for $\varphi^{u, in }_{\om}$) 
gives 
\be
\vert \alpha_{\om} \vert^2 +  \vert \tilde A_{\om} \vert^2 - \vert \beta_{-\om} \vert^2 = 1. \label{consnorm}
\ee 
In this expression, the minus sign comes from negative norm doublets, see \eqr{Dnorms}.

It should be noticed that the above enlarged Bogoliubov transformation governs the general case.
It applies indeed to any stationary situation when there is one type
of negative frequency modes (here the $u$-modes $\phi^u_{-\om}$). 
It should also be noticed that 
in situations with two sonic horizons, such as black hole white hole 
pairs~\cite{Corley:1998rk,Leonhardt:2008js,Steinhauer:2009aa}, 
the Bogoliubov  transformation will be more complicated than \eqr{Bog}.

\section{Fluxes, density fluctuations, and correlations}

Most of the work dedicated to the Hawking effect concentrates 
on the particle content, or the energy content, of the outgoing flux~\cite{Birrell+Davies,Wald:1995yp}. 
These two observables are related to $|\beta_\om|^2$ of \eqr{Bog}. 
However, it was also noticed that there exist Einstein-Podolski-Rosen (EPR) correlations between the outgoing particles 
and their partners with negative frequency $\om$. Unlike the energy flux, these correlations are weighted by $\alpha_\om \beta_\om^*$
and originate from interfering terms (i.e., nondiagonal in the occupation number). In the case of gravitational black holes, little attention has been given to the late time correlations because they are hidden for the external observers since the partners are trapped inside the horizon~\cite{Brout:1995rd}. 
Nevertheless these correlations have well-defined properties~\cite{Massar:1996tx}. Moreover, they extend to the past 
and can be revealed by sending quanta to stimulate the emission process~\cite{Wald:1976ka,Carlitz:1987aa,Carlitz:1987ab}.

With the advent of acoustic black holes, the situation 
completely changes because one has access to both regions,  
and can therefore probe the EPR correlations. In this respect, acoustic black holes are similar to what is found 
in the homogeneous time-dependent BEC~\cite{Barcelo:2003et,Fedichev:2003bv,Jain:2007gg,Weinfurtner:2008if}, 
and inflationary cosmology. In that case, the primordial fluctuations, 
seeds of the galaxy clusters and of the temperature anisotropies in the Cosmic Microwave Background, 
also result from pair creation~\cite{Mukhanov:1992me}. Moreover, 
their correlations possess a well-defined space-time structure~\cite{Campo:2003pa} which, 
on the one hand, affects today's observables because both members are within our Hubble patch,
and on the other hand, is very similar to that found in time-dependent 
BEC~\cite{Fedichev:2003bv,Carusotto:2009ca}.
 
We start our analysis with the 3 occupation numbers, with and without 
an initial temperature. The first case is similar to that of \cite{Macher:2009tw}, 
but many new features arise because of the different structure of 
the wave equation and because both $c$ and $v$ vary. 
The second case is completely new, as are the next sections where we relate 
the coefficients of \eqr{Bog} to both local and nonlocal density fluctuations. 

\subsection{Occupation numbers in the initial vacuum}

Let us first consider the in vacuum $| 0_{in} \rangle $, i.e., the state annihilated by the 
destruction operators $\hat a_\om^{u, in }$, $\hat a_\om^{v, in }$ and $\hat a_{-\om}^{u, in }$, 
defined by the in modes through 
\eqr{aoverl}. For gravitational black holes,
this is the physically relevant state 
(for outgoing $u$-configurations) 
after a few $e$-folding times $\Delta t=1/\kappa$. 
In fact, because of the exponential redshift effect 
associated with the near horizon propagation, see \eqr{nhr2}, 
the transient effects due to infalling quanta are exponentially rapidly washed out. 
Therefore, the choice of the initial distribution of quanta 
does not affect the stationary properties of the outgoing flux. 
This is no longer necessarily true for acoustic black holes, because the
dispersion limits the number of $e$-folding times during which \eqr{nhr2} applies~\cite{Balbinot:2006ua,Jacobson:2007jx}.
Thus, one should analyze each case to see if the vacuum state provides a reliable approximation. 

Assuming this is the case, the mean occupation numbers are 
\ba
\bar n^{\rm vac}_{\om}  &=&\langle 0_{in} | \hat a^{u, \, out\, \dagger}_\om \, \hat a^{u, \, out}_\om \, | 0_{in} \rangle 
= \vert \beta_{\om} \vert^2, 
\nonumber \\ 
\bar n^{{\rm vac}, \, v}_{\om} & = &  \langle 0_{in} | \hat a^{v, \, out\, \dagger}_\om \, \hat a^{v, \, out}_\om \, | 0_{in} \rangle 
=  \vert \tilde B_{\om} \vert^2, \label{threeoccn}\\
\bar n^{\rm vac}_{-\om} &=& \langle 0_{in} | \hat a^{u, \, out\, \dagger}_{-\om } \, \hat a^{u, \, out}_{-\om} \, | 0_{in} \rangle 
=
\vert \beta_{-\om} \vert^2 +    \vert  B_{\om} \vert^2 \nonumber\\
&=& \bar n^{{\rm vac}}_{\om} + \bar n^{{\rm vac}, \, v}_{\om} . \nonumber
\ea
These expressions follow when using \eqr{Bog} and \eqr{aoverl} to express the out
operators in terms of in ones.

When compared with the simpler 
case where $u$ and $v$ modes decouple~\cite{Brout:1995rd} 
the main novelty is that $v$-quanta are also produced (to the left of the horizon,
in the ``inside'' region when using the gravitational analogy). Their 
occupation number is $\bar n^{{\rm vac}, \, v}_{\om}$.
Because of this, $\bar n^{\rm vac}_{\om}$, the numbers of $u$ phonons emitted to the right 
is always smaller than that emitted to the left: 
$\bar n^{\rm vac}_{-\om}$. In fact, $\bar n^{\rm vac}_{-\om} = \bar n^{\rm vac}_{\om} +  \bar n^{{\rm vac}, \, v}_{\om}$ tells us that in addition to the usual ``Hawking'' channel, there is a new channel where one of the partner is a $v$-quantum. When the latter is negligible, for $\vert \tilde B_{\om} \vert^2 \ll \vert \beta_{\om} \vert^2$, one recovers the simpler case where $\bar n^{\rm vac}_{-\om} = \bar n^{\rm vac}_{\om}$. 
 
\subsection{Occupation numbers from an initial thermal state\label{Tini}}
 
In a BEC, in realistic situations, there will always be some residual temperature. The order of magnitude of this temperature is given by the chemical potential $\mu$ of Eq.~(\ref{tuned}), and thus inversely proportional to the healing length $\xi$. Moreover, the characteristic wavelength of the condensate inhomogeneity $\sim c/\kappa$ will generally be larger than $\xi$. Therefore, since Hawking temperature is $k_B T_H = \hbar \kappa/ 2 \pi$, the initial distribution of phonons could well hide the Hawking effect. 

For concreteness, to characterize the initial state, we assume that the three distributions have the same temperature. Using $\Om^{in}$, the initial value of the comoving frequency, the three initial occupation numbers $\bar n^{in}_\om, \bar n^{in,\,  v}_\om, \bar n^{in}_{-\om}$ are
\be
\bar n^{in, a}_\om = \left( e^{\beta_T \, \Om^{in, a}(\om)} - 1\right)^{-1} ,
 \label{thermal}
\ee
where $\beta_T$ is 
related to the (initial) temperature by $k_B T_{in} = \hbar /\beta_T$.
This choice means that, before the scattering in the near horizon region, the temperature
measured in the frame comoving with the fluid is the same for all modes. 

However, because of the scattering, modes sharing the same  
constant frequency $\om$ mix. Hence we must use $\om$ to characterize
the initial distributions. For each type of modes,
we thus need to express $\Om^{in}$ in terms of $\om$.
This explains the presence of the index $a$ in the comoving frequency 
in the above equation. 
Using \eqr{reldisp}, 
the values corresponding to the three in modes of \eqr{Bog} are
\be
\begin{split}
\Om^{in, u}(\om) &= \om - v_- \, k^u(\om), \\
\Om^{in, v}(\om) &= \om - v_+ \, k^v(\om), \\
\Om^{in, u}(-\om) &= - \om - v_- \, k^u(-\om),
\end{split}
\label{Omin}
\ee
where the three roots $k^u(\om) > 0, k^v(\om) < 0, k^u(-\om) > 0$, are clearly seen
in \figr{fig::reldispsols}.

Given the initial occupation numbers, \eqr{Bog} fixes the final ones to be
\ba
\bar n^{\rm fin}_{\om}  &=& \bar n^{in}_\om + 
 \vert A_{\om} \vert^2 \, (\bar n^{in, \, v}_\om - \bar n^{in}_\om) \nonumber \\
& & +  \vert \beta_{\om} \vert^2 \, (1+ \bar n^{in}_{-\om} + \bar n^{in}_\om), 
\nonumber \\ 
\bar n^{{\rm fin}, \, v}_{\om} & = &  \bar n^{in, \, v}_\om + 
 \vert \tilde  A_{\om} \vert^2 \, (\bar n^{in}_\om - \bar n^{in, \, v}_\om) \nonumber\\
  & & + \vert  \tilde B_{\om} \vert^2 \, (1+ \bar n^{in}_{-\om} + \bar n^{in, \, v}_\om), \label{betathreeoccn}\\
\bar n^{\rm fin}_{-\om} &=& \bar n^{in}_{-\om} + 
 \vert \beta_{-\om} \vert^2 \, (1+ \bar n^{in}_{-\om} + \bar n^{in}_\om) \nonumber \\
& & + \vert B_{\om} \vert^2 \, (1+ \bar n^{in}_{-\om} + \bar n^{in, \, v}_\om). \nonumber
\ea
The interpretation of these equations is clear. The first term is the corresponding initial occupation number. Then for the first two equations, 
the second term is due to the elastic scattering between $u$ and $v$ modes
which adds (or subtract) particles according to the strength of the scattering,
whereas the last term is due to the induced emission which involves both the 
partner's initial occupation number $\bar n^{in}_{-\om}$ and that of the species 
itself.
In the third line instead, 
one has two induced emission terms because there are two ``pair creation'' channels 
and no ``elastic'' channel.

\subsection{How to get rid of thermal noises?\label{discussion}}

From the first equation in \eqref{betathreeoccn}, one can easily imagine that the Hawking radiation (HR), i.e., the spontaneous creation of pairs weighted by  $ \vert \beta_{\om} \vert^2$ given in the first line of \eqr{threeoccn}, might be hidden by the presence of initial distributions. See also~\cite{Wuester:2008kh} for a discussion of the consequences of three-phonon interactions. 

There is yet another difficulty which can complicate the detection of HR, namely the possibility to distinguish right- from left-moving phonons. In the case one can,  detecting HR requires that $ \vert \beta_{\om} \vert^2$ be larger than, or at least of the same order as, both $\bar n^{in}_\om $ and   $\vert A_{\om} \vert^2 \, \bar n^{in, \, v}_\om$. The first condition could be satisfied because the initial distribution of $u$ modes can be significantly redshifted. 
By this we mean that one can have $\Om^{in, u}(\pm\om) \gg \om$ and thus possibly $\Om^{in,u}(\pm \om)/T_{in}\gg \om/T_H$, although {\it a priori} $T_{in}\gg T_H$. The second condition might be more problematic because the $v$ modes are hardly redshifted. Nevertheless, the inequality $\vert A_{\om} \vert^2 \, \bar n^{in, \, v}_\om \lesssim \vert \beta_{\om} \vert^2$ could also be satisfied because, as we shall see, in certain cases the $u-v$ mixing is very small. Therefore, when one is able to distinguish left- from  right-moving phonons, it could be possible to detect Hawking radiation, even when the initial temperature is larger than Hawking temperature. In Sec.~\ref{temperature}, we study numerically realistic situations and confirm this possibility. In the case one cannot distinguish left- from right-moving phonons, the situation is much worse. The dominant noise term would be $\bar n^{in, \, v}_\om$, and an initial temperature larger than Hawking temperature would hide Hawking radiation. 

Before proceeding to the numerical analysis, we study density fluctuations and nonlocal density correlations, firstly because phonon occupation numbers are not directly measurable (see however \cite{Schutzhold:2006aa}) whereas density fluctuations are, and secondly because nonlocal correlations are amplified by initial thermal distributions instead of being smeared  by them. 

The reader interested in the spectral properties can read Sec.~\ref{numres} first and go back afterwards to the next section.

\subsection{Density fluctuations}

Given that $\phi$ is a complex field, there are several ways to characterize the fluctuations in a BEC: either through the density correlation function which is governed by ${\rm Re}\,  \phi$, see \eqr{ReIm}, or through the phase correlations governed by ${\rm Im}\,  \phi$, or even through the crossed phase-density correlations. In what follows we only discuss the density-density correlations as the extension to the two other types is easily made. 

To simplify the forthcoming expressions we introduce the field operator $\hat \chi = \phi + \phi^\dagger$. In stationary cases, it can be decomposed as in \eqr{sumoverom}, and in terms of the same operators as those of \eqr{modeinom}. The only change is that the wave functions $\phi^a_\om,\, \varphi^a_\om$ are all replaced by 
\be
\chi^a_\om(x) = \phi^a_\om(x) + \varphi^a_\om(x). 
\ee
Since this correspondence applies to both the in and out sets, and since both $\phi^a_\om$ and $ \varphi^a_\om$ obey transformation of \eqr{Bog}, 
the three initial $\chi^{in, \, a}_\om$ are also related to the three final $\chi^{out, \, a}_\om$ by \eqr{Bog}. Therefore, even though $\hat \chi$ is not a canonical field, as it does not obey canonical commutators, it can be treated as a genuine quantum field when computing correlation functions.

The statistical properties of the density fluctuations encoded in a given state are characterized by the anti-commutator
\ba
G^{\rm in}(t,x;t', x') &= & \frac{1}{2}{\rm Tr}[\,  \hat \rho^{\rm in} 
 \left\{\hat \chi(t,x), \, \hat \chi(t',x')\right\}  ]
\nonumber \\
&= & \int_{-\infty}^\infty d\om \, e^{-i \om (t- t')} \, 
G^{\rm in}_\om(x,x') ,
\label{Gac}
\ea
where $\hat \rho^{\rm in}$ is the initial density matrix, since
we  work in the Heisenberg representation. 
In the second line we passed to a Fourier transform since we assumed that 
both the condensate and the state are stationary (in the ``preferred'' frame).

When the initial state $\hat \rho^{\rm in}$ is incoherent, 
using the in basis to express $\chi$, only three terms  having the same structure are obtained
\be
\begin{split}
G^{\rm in}_\om(x,x') = \left( \bar n^{{\rm in}}_\om + 1/2 \right) \, \chi^{in, \, u}_\om(x) \left[ \chi^{in, \, u }_\om(x') \right]^* \\
 +  \left(\bar n^{{\rm in}, \, v}_\om + 1/2 \right) \, \chi^{in, \, v}_\om(x) \left[ \chi^{in, \, v}_\om(x') \right]^* \\
+ \left(\bar n^{{\rm in}}_{-\om} + 1/2 \right)\, \left[\chi^{in, \, u}_{-\om}(x)\right]^* \chi^{in, \, v }_{-\om}(x').
\end{split}
\label{Ginin}
\ee
This expression is valid for $\om > 0$; for $\om < 0$, 
one has $G^{\rm in}_{-\om}(x,x') = [G^{\rm in}_{\om}(x,x')]^*$
since $G^{\rm in}(t,x;t',x')$ is real. In the above equation, the initial occupation numbers are given by
\be
n^{{\rm in}, \, i}_\om \times \delta^{ij} =  {\rm Tr}[\hat \rho^{\rm in}\, a_\om^{in,\,  i\, \dagger}\,   a_\om^{in, \, j} ].
\label{inpowers}
\ee
Due to the incoherence of $\hat \rho^{\rm in}$, only the diagonal terms remain.
Additional terms would be obtained if the state $\hat \rho^{\rm in}$ 
contained correlations among the 
initial configurations. In what follows we assume it does not (see however App.~\ref{coher}).

Because of the scattering near the sonic horizon, 
$G^{\rm in}_\om(x,x')$ has a rather complicated structure, 
as can be seen by decomposing the in modes
into out ones. In fact $G^{\rm in}_\om(x,x')$ 
encodes both local observables related to the occupation numbers of \eqr{betathreeoccn},
and non-local ones governed by correlators such as 
${\rm Tr}[\hat \rho^{\rm in}\, a_\om^{out,\,  i}  a_\om^{out, \, j} ]$ 
with $i \neq j$. 

\subsection{Coincidence point limit \label{densfluctth}}

Far from the horizon so that the scattering is completed, i.e., when $|x| \gg D c_0/\kappa$, 
one should express the in modes in terms of their 
asymptotic plane wave content $\sim e^{i k^a_\om x}$. 
To ease the reading, we shall call these asymptotic contributions
by the corresponding wave, $\chi_\om^{in, \, a}$ or $\chi_\om^{out, \, a}$,
for which the amplitude of this contribution is unity, and, to avoid any misinterpretation,
we shall add an upper index ``as'' to make clear 
that only the unit, plane wave contribution should be kept. 
In terms of these, using \eqr{betathreeoccn}, in the asymptotic right region, one gets
\be
\begin{split}
G^{\rm in}_\om(x,x) \to 
( \bar n^{{\rm fin}, \, u}_\om + 1/2) \times |\chi_\om^{out, \, u, \, \rm as}|^2 \\
+\, (\bar n^{{\rm in}, \, v}_\om + 1/2)\times |\chi_\om^{in, \, v, \, \rm as}|^2 ,
\end{split}
\label{n1}\ee
whereas, on the left, the ``power'' is asymptotically equal to
\be
\begin{split}
G^{\rm in}_\om(x,x) \to (\bar n^{{\rm fin}, \, v}_\om + 1/2) \times |\chi_\om^{out, \, v, \, \rm as}|^2 \\
+\, (\bar n^{{\rm fin}, \, u}_{-\om} + 1/2) \times |\chi_{-\om}^{out, \, u, \, \rm as}|^2 \\
+\, (\bar n^{{\rm in}, \, u}_\om + 1/2)\times |\chi_\om^{in, \, u, \, \rm as}|^2 \\
+\, (\bar n^{{\rm in}, \, u}_{-\om} + 1/2) \times  |\chi_{-\om}^{in, \, u, \, \rm as}|^2 .
\end{split}
\label{n2}
\ee
Since the point $x=x'$ lives in an asymptotic region where the condensate is
homogeneous, the norm of the asymptotic modes $\chi^{\rm as}_\om$ 
is 
$x$ independent. 
Notice also that we discarded all oscillatory terms, such as $\chi_\om^{out, \, u, \, \rm as}
(\chi_\om^{in, \, v, \, \rm as})^* \sim e^{i(k^u_\om - k^v_\om)x}$, 
because they rapidly oscillate as $x \to \infty$, 
and thus drop out when averaging over $\om$.

The interpretation of the above equations is clear. 
When working in the state $\hat \rho^{\rm in}$, 
the two 
asymptotic values of the correlator are the sum of 
the contributions of the waves that have been scattered, governed 
by the final occupation numbers, 
\be
n^{{\rm fin}, \, i}_\om =  {\rm Tr}[\hat \rho^{\rm in}\, a_\om^{out,\,  i\, \dagger}\,   a_\om^{out, \, i} ],
\label{finpowers}
\ee
and of the waves which have not propagated through the horizon region, 
and whose occupation numbers are the initial ones given in \eqr{inpowers}. 
From \eqr{n1} one clearly sees that the second term (the $v$ contribution) 
will hide the Hawking radiation
whenever $\bar n^{{\rm in}, \, v} $ 
is much larger than $\bar n^{{\rm fin}, \, u}$, 
which is the case in realistic situations, as we shall see in Sec.~\ref{numresdensfluct}. 

It is therefore also of interest to compute 
\begin{align}
{\cal F}(t,x;t',x') &= \frac{i}{2\rho_0} (\partial_{x'}-\partial_x){\rm Tr}
\left[\hat\rho^{\rm in} \, \Psi^\dagger(t',x') \, \Psi(t,x)\right]
\nonumber
\\
 &= \int_{-\infty}^{\infty} d\om \, e^{-i\om(t-t')} \, {\cal F}_\om(x,x').
\end{align}
In the coincidence point limit $(t,x)=(t',x')$, $\hbar \rho_0{\cal F}/m$ is the atom flux at $(t,x)$. 
It is the sum of $\rho_0v=\rho_0 \hbar k_0/m$, the flux of the condensed atoms, and of the integral over $\om$ of ${\cal F}_\om(x,x)$.
In the asymptotic right region, using \eqr{statio}, for $\om>0$, one has ${\cal F}_\om = {\cal F}_\om^{\rm dr} + {\cal F}_\om^{\rm com}$ where
\be
\begin{split}
&{\cal F}_\om^{\rm dr} = k_0 \times \left[\bar n^{\rm fin}_\om \,  |\phi^{u,out,\rm as}_\om(x)|^2 + \bar n^{in,v}_\om \, |\phi^{v,in,\rm as}_\om(x)|^2\right],\\
&{\cal F}_\om^{\rm com} = \bar n^{\rm fin}_\om \,  k^{u,out}_\om |\phi^{u,out,\rm as}_\om(x)|^2 + \bar n^{in,v}_\om \,  k^{v,in}_\om |\phi^{v,in,\rm as}_\om(x)|^2.
\end{split}
\label{flusso}
\ee
For $\om<0$, one has the same expressions with opposite signs, with $\phi_\om$ replaced by $\varphi_\om$, and $\bar n_\om$ replaced by $\bar n_\om +1$. ${\cal F}_\om^{\rm dr}$ arises from the uncondensed atoms and is due to the dragging of the background, while ${\cal F}_\om^{\rm com}$ is the atom flux measured in the frame comoving with the condensate. As in the case of the density fluctuations, the term arising from the initial distribution of $v$-phonons largely dominates. However, since $k^{v,in}_\om<0$, ${\cal F}^{\rm com}_\om$ 
is related to the \emph{difference} of the terms appearing in $G^{\rm in}_\om(x,x)$, up to $k$-dependent factors (different for each term). Thus, if one has access to both the atom flux and the density fluctuations in the right asymptotic region, there is a greater hope to have access to $\bar n^{\rm fin}_\om$. 

Note that any other local observable constructed out of two fields, like the depletion, will have the same structure in the right asymptotic region and will thus suffer from the same limitations, namely it will be largely dominated by the contribution of $n^{in,v}_\om$. This reinforces the interest to consider nonlocal observables.

\subsection{Long distance correlations.}

\subsubsection{Late time entanglement\label{latetiment}}

We now study the long distance correlations, for $|x - x'| \gg  D c_0/\kappa > \xi_0$. In this case $G^{\rm in}_\om(x,x')$
displays a rich structure. Considering $x > 0$, $x'< 0$, one gets {\it a priori} eight terms, since one has two asymptotic modes $\chi^{\rm as}$ on the right, and four on the left, see \eqr{modeinom2}. However, many of them drop out because they destructively interfere upon integrating over $\om$, see the discussion after \eqr{n2}. Hence, when the initial state $\hat \rho^{\rm in}$ is incoherent, no (long distance) correlations among in modes exist. Similarly, all terms mixing in and out $\chi^{\rm as}$ modes will  destructively interfere. In fact, only out phonons are entangled by the scattering in the near horizon region. Hence only correlations among asymptotic out modes will contribute. Given that in the subsonic region, there is only one asymptotic out mode, $\chi_\om^{out,\, u, \, \rm as}$, and in the supersonic region, two such modes exist, for $x > 0$, $x'< 0$, one has 
\be
\begin{split}
G^{\rm in}_\om(x,x') = \chi_\om^{out, \, u, \, \rm as}(x)  \times \Big\{ {\cal A}_\om \left[\chi_\om^{out, \, v,\,  \rm as}(x')\right]^* \\
 +\, {\cal B}_\om \,  \chi_{-\om}^{out, \, u, \, \rm as}(x') \Big\}.
\end{split}
\label{ldcorr1}
\ee
When both $x$ and $x'$ are taken in the subsonic region, no long distance correlations can develop
since $\chi_\om^{out,\, u, \, \rm as}$ is the only asymptotic out mode. On the contrary, when both $x$ and $x'$ are negative, in the supersonic region,
the two asymptotic modes are entangled and this will show up in
\ba
G^{\rm in}_\om(x,x') &=& {\cal C}_\om \, 
\chi_\om^{out, \, v, \, \rm as}(x) \, 
 \chi_{-\om}^{out, \, u, \, \rm as}(x'). 
\label{ldcorr2}\ea
Given the $3 \times 3$ character of \eqr{Bog}, three types of late time correlations could have been expected since three different couples of out modes can be formed. 

To compute ${\cal A}_\om,\,  {\cal B}_\om$ and $ {\cal C}_\om$ one can either express the r.h.s. of \eqr{Ginin} in terms of out modes and identify
the coefficients multipying the corresponding couple of $\chi_\om^{out, \rm as}$, or equally start with \eqr{Gac} and decompose the field $\chi$ using the out basis. Adopting the second, more rapid method, we get 
\be
\begin{split}
{\cal A}_\om &= {\rm Tr}[\hat \rho^{\rm in}\, a_\om^{out,\,  u} 
\,   a_\om^{out, \, v\, \dagger} ],\\
 {\cal B}_\om &= {\rm Tr}[\hat \rho^{\rm in}\, a_\om^{out,\,  u} \, a_{-\om}^{out, \, u}] ,\\
 {\cal C}_\om &= {\rm Tr}[\hat \rho^{\rm in}\, a_\om^{out,\,  v} \,   a_{-\om}^{out, \, u} ] ,
\end{split}
\label{ABC2}
\ee
where we have used the fact that the $a^{out}$ operators commute in each product. 
Unlike the coincidence point limit of $G^{\rm in}_\om(x,x')$ which is governed by diagonal
terms, see Eqs.~(\ref{n1}, \ref{n2}), 
the long distance correlations are governed by terms which are nondiagonal
in occupation number. This is exactly as in homogeneous situations~\cite{Campo:2003pa,Carusotto:2009ca} 
and in fact will always be found when the parametric amplification (or the scattering) conserves a quantity,
the frequency $\om$ here, the spatial wave-vector $\bf k$ in homogeneous cases. A straightforward calculation gives
\begin{align}
&{\cal A}_\om = \bar n^{in}_\om \, \alpha_\om \,  \tilde A_\om^* + 
n^{in, \, v}_\om \, A_\om\, \alpha_\om^{v\, *}  + 
(\bar n^{in}_{-\om} +1)\, \beta_\om^* \, \tilde B_\om,
\nonumber \\
&{\cal B}_\om = \bar n^{in}_\om \, \alpha_\om \, \beta_{-\om}^* + 
\bar n^{in, \, v}_\om  \, A_\om \, B_\om^{*}+ 
 (\bar n^{in}_{-\om} +1)\, \beta_\om^* \, \alpha_{-\om}, \label{ABCs}\\
&{\cal C}_\om = \bar n^{in}_\om \, \tilde  A_\om \, \tilde \beta_{-\om}^* + 
\bar n^{in, \, v}_\om \, \alpha_\om^v \,  B_\om^{*} + 
(\bar n^{in}_{-\om} + 1)\,  \tilde B_\om^* \, \alpha_{-\om}.\nonumber
\end{align}
In the vacuum, the residual correlations all involve the negative frequency modes
because these enter in both pair creation channels.

At this point, an important observation should be made. 
Because the initial distributions $\bar n^{in, \, a}$ 
only appear in factors multiplying terms already present 
in the in vacuum, 
the long distance correlations will not be erased by the presence of initial quanta. 
In fact an initial temperature will in general {amplify} 
the long distance correlations induced by interactions.~\footnote{We 
added ``in general'' because the terms in the r.h.s. of \eqr{ABCs} have no definite sign. 
Hence, an increase in the temperature could lower, in some (rare) cases, the resulting amplitudes of the
l.h.s. of \eqr{ABCs}. It is interesting to point out that when the $u-v$ mixing is small, i.e. when the $A$ and $B$ 
coefficients can be set to zero, 
only two terms in $\cal B$ remain (the first and the third one),
and that these will always add up, thanks to the unitarity of the reduced $2 \times 2$ Bogoliubov transformation.} 
It was numerically observed in \cite{Carusotto:2008ep}, as was the fact that the ${\cal A}_\om$ coefficient, arising from the product $a_\om^{ u} a^{v \dagger}_\om$, is nonzero even in a subsonic flow. \footnote{Because of the orthogonality of the scattering matrix of \eqr{RT} it would be zero unless $n_\om^{in} \neq n^{in, \, v}_\om$. We thank C. Mayoral and A. Fabbri for drawing our attention to this particular point.}

We finally remind the reader that the above correlations were obtained using the BdG equation (\ref{centraleq0}) which neglects phonon interactions~\cite{Wuester:2008kh}. One might therefore worry that the entanglement is reduced upon taking into account such interactions. We refer to Refs.~\cite{Campo:2005sy,Campo:2008ju,Campo:2008ij} for an analysis of this point in a cosmological context. In brief, the weakness of the nonlinearities
guarantees that the entanglement is hardly reduced.

\subsubsection{Spatial structure of long distance correlations\label{spatialstruct}}

At fixed $\om$ no spatial structure emerges from Eqs.~(\ref{ldcorr1}, \ref{ldcorr2}). 
To get the spatial properties of the correlations, one needs to take the inverse Fourier transform. Then, as it is the case when considering wave-packets, 
 constructive interferences will develop along the characteristics of the mode equation. 
To ease the reading of the forthcoming expression we found convenient to return to \eqr{Gac}
and to re-introduce $t$ and $t'$. 

As usual, the constructive interference condition gives the stationary phase condition $\partial_\om S= 0$.
In the present case, using a WKB approximation for the modes $\chi_\om^a(x) \sim \exp(i \int^x dy k^a_\om(y))$,
the three phases of the terms weighted by ${\cal A}_\om ,\, {\cal B}_\om$ and ${\cal C}_\om$ are
respectively
\begin{align}
&\begin{split}S_A(t,t',x,x'; \om) = - \om (t-t') + \int^x_z dy k^u_\om(y) \\- \int^{x'}_z dy k^v_\om(y) + \arg( \ln {\cal A}_\om), \end{split}
\nonumber \\
&\begin{split} S_B(t,t',x,x'; \om) =  - \om (t-t') + \int^x_z dy k^u_\om(y) \\+ \int^{x'}_z dy k^u_{-\om}(y) + \arg( \ln {\cal B}_\om),\end{split}
\label{3St} \\
&\begin{split} S_C(t,t',x,x'; \om) =  - \om (t-t') + \int^x_z dy k^v_\om(y) \\+ \int^{x'}_z dy k^u_{-\om}(y) + \arg( \ln {\cal C}_\om),\end{split}
\nonumber
\end{align}
where $z$ is an arbitrary location where the absolute (unobservable) 
phase of the $\chi_\om^a$ is fixed.
The stationary phase condition gives
\be
\begin{split}
(t-t') =  \int^x_z dy \, \partial_\om k^u_\om(y) - \int^{x'}_z dy\,  \partial_\om k^v_\om(y) \\
+ 
\partial_\om \arg( \ln {\cal A}_\om), 
\end{split}
\label{phA}
\ee
for the first line, and similar equations for the second and third lines. 
It seems {\it a priori} that the choice of $z$ matters. However this is not the case because $z$
 enters in ${\cal A}_\om$ in such a way that a change of $z$ leaves the r.h.s. of the equation unchanged. 
(This is because the coefficients of \eqr{ABC2} 
contain the operators $a^\dagger, a$ which are ``contravariant'' with respect to a phase shift of the corresponding mode.) 
The physically meaningful phase that comes out of these expressions has the role of fixing 
the location where the interactions occur.~\cite{Campo:2003gb}

This result becomes exact in the limit where $x$ and $x'$ 
are taken far away from the scattering zone, and it amounts to put $z=0$ and to 
to drop the last term in \eqr{phA}, and similarly for the equations involving  $\cal B_\om$ and $\cal C_\om$. 
There could be some finite phase shift with respect to these WKB estimates, 
but these do not change with $x$ and $x'$, and hence give subdominant effects in the large $x$ limit.
In this limit, the properties of the correlation pattern derived from \eqr{phA} 
are thus independent of both the norm and the phase of the coefficient $\cal A_\om$. 
Hence the same long distance space-time pattern will be found both in the limit $\kappa \to 0$ and $\kappa \to \infty$,
i.e. in regimes which do not give the standard Hawking radiation. Therefore if Hawking radiation 
implies the pattern, the converse is not true (when defining Hawking radiation as the near thermal 
radiation associated with a finite surface gravity $\kappa$).
 
In the large distance limit, when putting $t=t'$, the stationary phase condition applied to \eqr{3St} gives, for the ${\cal A}_\om, {\cal B}_\om$ and ${\cal C}_\om$
terms respectively,
 \ba
\Delta t^{HJ, \, u}_\om(x) &=&  
\Delta t^{HJ, \, v }_\om(x'), 
\nonumber \\
\Delta t^{HJ, \, u}_\om(x) &=&  
\Delta t^{HJ, \, u}_{-\om}(x') , 
\nonumber \\
\Delta t^{HJ, \, v}_\om(x) &=&  
\Delta t^{HJ, \, u}_{-\om}(x') ,
\label{3HJt}
\ea
where 
\be
\Delta t^{HJ, \, a}_\om(x) =  \int^x_0 dy \, \partial_\om k^a_\om(y)
\label{HJta}
\ee is the  time it takes the $a$-type phonon
of frequency $\om$ to propagate from $x=0$ to 
$x$, 
in virtue of the  Hamilton-Jacobi 
equation determining the group velocity 
\be
v^a_{gr}(\om) = (\partial_\om \, k^a_\om)^{-1}.
\label{grv}
\ee
For each type of correlations, we see that 
the locus of constructive interference of, say, $x'$ given $x$,
is given by the value of $x'$ reached at the same lapse time it takes the 
partner to reach $x$, both phonons starting their journey near the horizon $x= 0$. 
In the large $x$ limit, the dominant contribution comes from the uniform motion 
outside the horizon region. Thus the above three constructive interferences 
occur at locations $x,x'$ related by
\be
\frac{x}{v_{gr}^a(\om)} = \frac{x'}{v_{gr}^b(\om)}.
\ee
Using \eqr{reldisp}, the asymptotic group velocities are, for $x \to + \infty$,
\ba
v_{gr}^u(\om) &=& \partial_k \Omega_+ + v_+ = c_+^2 k \, \frac{1+ k^2 \xi^2_+}{\Omega_+(k)} + v_+ ,
\label{vgr1}
\ea
for the right movers in the subsonic region, the ``Hawking quanta''; and, for $x \to - \infty$, 
\ba
v_{gr}^v(\om) &=& - \partial_k \Omega_- + v_- = - c_-^2 k \,  \frac{1+ k^2 \xi^2_-}{\Omega_-(k)} + v_-,
\nonumber \\
v_{gr}^u(-\om) &=& \partial_k \Omega_- + v_- = c_-^2 k \,  \frac{1+ k^2 \xi^2_-}{\Omega_-(k)} + v_-,
\label{vgr2}
\ea
for the $v$ quanta and the $u$ partners in the supersonic region.
In the dispersionless regime, for $k \xi_\pm\ll 1$, 
these group velocities are independent of $\om$ and respectively equal to 
$c_+ + v_+$ in \eqr{vgr1}, and $- c_- + v_-$, $c_- +v_-$ for \eqr{vgr2}. 
Therefore, in this regime, the correlation pattern is independent
of $\om$, and will show up in $G^{\rm in}(t,x;t'=t, x')$ of \eqr{Gac}.

These three types of correlations have been (numerically) observed in \cite{Carusotto:2008ep}.
They have been also correctly interpreted save for two aspects.
First, the  $\cal C_\om$  branch has been 
attributed to the ``partial elastic scattering'' of right movers of positive frequency.
If this explanation applies to the first term, it does not to the last two which involve pair creation 
$B_\om$ coefficients mixing $\chi^v_\om$ and  $\chi^u_{-\om}$. Second, the $\cal A_\om$ is claimed to ``originate from thermal effects''.
From the first line of \eqr{ABCs} we see that it is indeed 
amplified by thermal effects, but, because of the third term, 
it is already present in the vacuum. Being quadratic in frequency mixing coefficients, 
it is too weak to be easily seen in the numerical simulations. 
However using a black and white print of the arXiv version of \cite{Carusotto:2008ep},
these correlations are quite visible in the last two plots in Figure 2, see Figure 6 
for their orientation in the $x,x'$ plane. Upon contacting the authors, they agree that
these correlations are indeed present at zero temperature. When preparing the revised version of this paper, we became aware of \cite{Recati:2009aa}
which agrees with this point, as with the rest of our analysis,
and which also contains additional interesting results.

Concerning the relative amplitude of $\cal A_\om$, $\cal B_\om$ and $\cal C_\om$ it should be noticed that, in general, there is no clear ordering. 
Instead, when the $u-v$ mixing is weak, the Bogoliubov coefficients $A_\om$, $B_\om$ are 
much smaller than $\beta_\om$ (see Sec.~\ref{numresnv}) and therefore, $\cal B_\om$ is the largest. 
In this regime, one recovers the properties 
of the $\cal B_\om$ branch relating the $u$ modes across the horizon 
that have been known for a while. 
In the context of gravitational black holes, using relativistic fields, they 
can be found in \cite{Massar:1996tx,Brout:1995rd}.
As of acoustic black holes, it was understood in \cite{Brout:1995wp} 
that the late time behavior of these correlations is essentially unaffected 
by dispersive effects,\footnote{In fact the correlation pattern is as robust as the occupation number $n_\om = \vert \beta_\om \vert^2$ 
and the stationarity 
against introducing dispersion. 
This is because $z_\om =\beta_\om/\alpha_\om$,
 the complex squeezing parameter which determines the final quantum state~\cite{Brout:1995rd}, 
is hardly affected. Its norm determines $n_\om$ 
whereas its phase determines the correlation pattern. 
The density (in $\om$) of modes subject to the Bogoliubov transformation ensures the stationarity.} see also \cite{Balbinot:2006ua,Jacobson:2007jx}.
Finally, that the $\cal B_\om$ correlations 
determine (in the hydrodynamical limit) 
the long distance density correlations in BECs was stressed in~\cite{Balbinot:2007de}.

In App.~\ref{coher} we present an 
alternative and simple way to characterize the correlation pattern. It consists in sending classical waves --described by highly excited coherent states--
towards the horizon. This approach is worth considering because it
allows to relate the above study of \eqr{Gac} both to the experiments in hydrodynamics described in \cite{Rousseau:2008aa},
and to the wave packet analysis of \cite{Brout:1995wp}.

\section{Spectral properties: numerical results\label{numres}}

\subsection{Numerical procedure}

The numerical procedure we used was elaborated from that of \cite{Macher:2009tw}. 
The wave equation of that reference was 
replaced by the BdG equation \eqr{centraleq0}, where the velocity $v$ and the sound speed $c$ 
both vary, see  \eqr{vdD}. The extraction of the Bogoliubov coefficients from the numerical solutions to \eqr{centraleq0} 
took into account the specific normalization of the modes $(\phi_\om, \varphi_\om)$. For a presentation of the procedure itself, we refer to  \cite{Macher:2009tw}.

\subsection{Value of the parameters\label{numbers}}

Let us determine the number of free parameters and their realistic ranges.\footnote{We are grateful to Eric Cornell for providing us with realistic experimental values for all parameters and to Jeff Steinhauer for further comments.} A typical value for the average sound speed $c_0$ is
\be
c_0 = 
0.15\,{\rm cm} \, \cdot {\rm s}^{-1}.
\ee
Assuming that the condensate is made out of $ ^{85}$Rb, the mass of the atoms is 
\be
m=1.5\times 10^{-25}\,{\rm kg}.
\ee
This yields the healing length, 
\be
\xi_0 = \frac{\hbar}{\sqrt{2}mc_0} = \frac{\sqrt{2}c_0}{\Lambda}\simeq 3.3\times 10^{-5}\,{\rm cm} .
\label{xiLambda}
\ee

The distance over which the variation of the sound speed and flow velocity takes place, that is, the distance separating the asymptotic regions where these speeds are constant, cannot be smaller than a few healing lengths. To reduce the number of free parameters, we assume in this work that it is of the order of $10\xi_0$ (see however Sec.~\ref{dispersive}). Then, given our parameterization \eqr{vdparam}, the gradient at the horizon is
\be
\kappa = \frac{c_0D}{5\xi_0}.\label{xiDkappa}
\ee

With \eqr{xiLambda}, this yields a relationship between $D$ and $\lambda=\Lambda/\kappa$:
\be
\lambda \simeq \frac{7}{D}.\label{constrLD}
\ee
as well as an expression for the Hawking temperature:
\be
T_H = \frac{\hbar\kappa}{2\pi k_B} \simeq D\times 1.1\,{\rm nK},
\ee
where we have used \eqr{xiDkappa} and the numerical values above. We 
choose as free parameters $D$ and $q$. Thus \eqr{constrLD} fixes $\lambda$.

Fairly large relative variations around the average sound speed $c_0$ can be achieved experimentally, for instance by means of a Feshbach resonance that modifies the coupling constant $g$ along the flowing BEC. Hence we shall consider values of $D$ from $0.1$ to $0.7$ (higher values are not excluded experimentally, but proved difficult to reach with our code). The corresponding values of $\lambda$ go from $70$ to $10$, while the order of magnitude of Hawking temperature is
\be
T_H\simeq 0.1-0.8\,{\rm nK}.
\ee
In general there will be a variation of both the sound speed and the flow velocity. Which one varies most depends on the experimental conditions. 
In~\cite{Carusotto:2008ep}, $q$ was taken to be zero. 
In the following, the typical value of $q$ will be  $ 0.3$. In Sec.~\ref{effectinh}, we shall nevertheless explore the whole range $0<q < 1$.

In realistic conditions a condensate has an effective temperature approximately fixed by the chemical potential $\mu \simeq m c^2_0$. To be able to quantitatively compare Hawking temperature with this effect, 
we define 
\be
T_{\xi_0} = \frac{m c^2_0}{k_B} = \frac{\hbar c_0}{\xi_0k_B}\simeq 30\,{\rm nK},\label{Thealing}
\ee
and in the following we consider initial temperatures (entering \eqr{thermal})
\be
T_{in}=\tau \, T_{\xi_0}, 
\label{tau}
\ee
with $\tau$ ranging from $1/3$ to $1$, so that $T_{in}$ goes from $10\,{\rm nK}$ to $30\,{\rm nK}$. Equation~\eqref{Thealing} together with~\eqref{xiLambda} yields
\be
\frac{T_{in}}{T_H} = \tau\times\sqrt{2}\pi\lambda.\label{TresTHL}
\ee
If $\lambda$ is constrained by \eqr{constrLD}, this can be rewritten as
\be
\frac{T_{in}}{T_H}\simeq 30\frac{\tau}{D}.\label{TresTHD}
\ee
The fact that $T_{in}$ is about two orders of magnitude higher than the Hawking temperature 
will surely complicate 
measuring the spectral properties of the Hawking radiation. 

In the following sections, we start by studying these spectral properties assuming the condensate has zero temperature. 
The analysis when there is an initial temperature is then performed in Sec.~\ref{temperature}. Our main goal is to understand how the adimensional parameters $D,\lambda,q$ affect the fluxes. To this end it proves very convenient to work with the rescaled frequency $\om/\kappa$ and rescaled energy flux $f_\om$ defined below. 

\subsection{Spectral properties of HR at zero temperature\label{zerotemp}}

Let us first study the energy flux of positive frequency $u$ quanta 
on the right of the horizon, i.e., 
what corresponds to Hawking radiation. 
We denote by $F$ the total energy flux, 
and we define the energy flux density as:
\be
f_\om = \frac{2\pi}{k_B T_H}\frac{dF}{d\om}
= 2\pi\frac{\om}{\kappa}|\beta_\om|^2.
\ee
The factor $2\pi/k_BT_H$ is here for convenience, so that $f_\om$ be dimensionless and normalized to $1$ at $\om=0$ when the occupation number $n_\om = |\beta_\om|^2$ is the standard, Planckian one with temperature $T_H$. 
To characterize the deviations from the standard flux, it is also convenient 
to use 
the effective temperature $T_\om$ defined as
\be
n_\om = \frac{1}{\exp(\hbar \om/k_BT_\om)-1}. 
\ee

\subsubsection{Typical spectra\label{typicalspectra}}
\begin{figure*}
\includegraphics{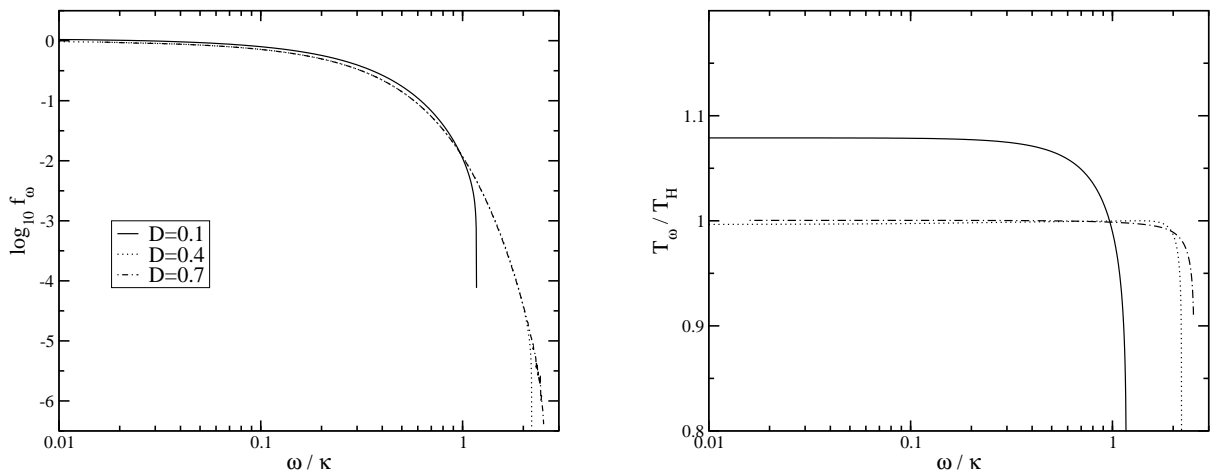}
\caption{Energy flux density $f_\om$ (left plot) and effective temperature $T_\om/T_H$ (right plot) versus $\om/\kappa$, for $(D,\lambda)=(0.1,70)$, $(0.4, 18)$ and $(0.7,10)$. $q$ is fixed to $0.3$.\label{fig::typicalfom}}
\end{figure*}

In \figr{fig::typicalfom} $f_\om$  and $T_\om/T_H$ are represented versus $\om/\kappa$ for 
$D=0.1$, $0.4$ and $0.7$. The corresponding $\lambda$ are fixed by \eqr{constrLD} and are respectively $\lambda=70$, $18$ and $10$. 
The parameter $q$ specifying the relative contribution of $c$ and $v$ to the gradient at the horizon, see \eqr{vdD}, is fixed to $0.3$. 
The three values of $\om_{\rm max}/\kappa$ are respectively $1.17$, $2.2$ and $2.54$. 
The energy flux and $T_\om$ quickly drop to zero when approaching $\om_{\rm max}$, as expected. Until short before $\om_{\rm max}$, $T_\om$ is nearly constant. 

Note nevertheless  that this constant temperature can differ 
from $T_H$: for $D=0.1$, the asymptotic temperature when $\om\to 0$ is equal to $T_0 = 1.08 \, T_H$. For $D=0.4$ and $D=0.7$, we found $T_0=0.996 T_H$ and $T_0=1.0004 T_H$ respectively, so $T_0$ differs from $T_H$ only by a fraction of a percent when $D$ is large enough. 
Note also that the scale separation condition $\lambda\gg 1$ is not the relevant criterion to predict the importance of the deviation with respect to the standard spectrum, since $\lambda$ is much larger for $D=0.1$ than for the other two values. Instead, it is the ratio $\om_{\rm max}/\kappa$ 
that controls the deviation from the standard temperature. This is confirmed in Sec.~\ref{robustness}.

The thermality of the spectra can be characterized more precisely. To this end we define
\be
\Delta_{T_0} = \frac{T_0-T_{\om=k_BT_0/\hbar}}{T_0}, 
\ee
that measures the running of the temperature. 
For the three spectra of \figr{fig::typicalfom}, we found $|\Delta_{T_0}|\simeq 0.1\%$. 
This establishes that, to a very good approximation, 
the energy spectra are Planckian, but truncated very close to $\om_{\rm max}$. 
At this point it should be stressed that to reach this result, nowhere
have we used the gravitational analogy. In fact the analogy is validated by our results since it would have predicted a thermal spectrum at
the standard Hawking temperature, missing however the truncation near $\om_{\rm max}$ and the small running $|\Delta_{T_0}|$).

\subsubsection{$\om_{\rm max}/\kappa$ controls the deviation from the standard spectrum\label{robustness}}

\begin{figure}
\includegraphics{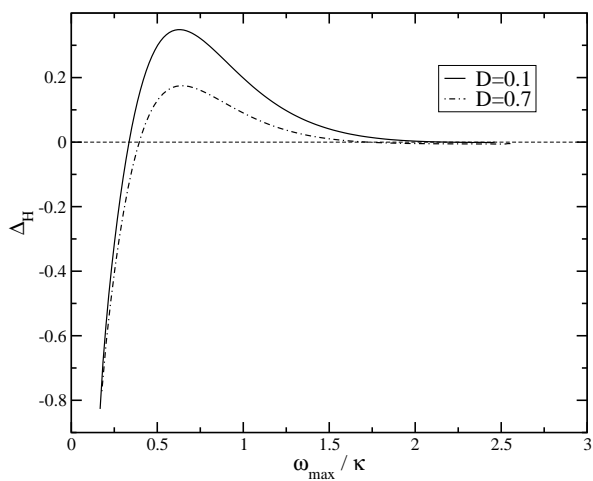}
\caption{$\Delta_H$ versus $\om_{\rm max}/\kappa$ for $D=0.1$ and $D=0.7$. $q$ is fixed to $0.3$.\label{fig::robustness}}
\end{figure}

In this section, in order to show that $\om_{\rm max}/\kappa$ is the main quantity that controls the modifications 
with respect to the standard Planckian spectrum with temperature $T_H$, we relax the constraint \eqref{constrLD}, and allow for arbitrary values of $\lambda$ for a given $D$. 
We characterize the deviation with respect to the standard result by 
the relative difference 
\be
\Delta_H = \left.\frac{f_\om - f^H_\om}{f^H_\om}\right|_{\om= \om_H}, 
\ee
as a function of $\om_{\rm max}/\kappa$, 
where $f^H_\om$ denotes the Planckian energy flux density with temperature $T_H$ 
and where $\om_H = k_BT_H/\hbar$. 

In \figr{fig::robustness}, for $q = 0.3$, the curves associated with
$D=0.1$ and $D=0.7$ have similar shapes, with a maximum around $\om_{\rm max}/\kappa=0.6$, 
with height $0.35$ for $D=0.1$ and $0.18$ for $D=0.7$. The deviation then decreases and in both cases becomes less than a percent as soon as $\om_{\rm max}/\kappa>2$. This confirms what was found (in a different setup) in Ref.~\cite{Macher:2009tw}. This is not the end of the story however, since as we now show, there can be significant deviations from thermality in a BEC, even when $\om_{\rm max}/\kappa$ is large, depending on the value of the parameter $q$.

\subsubsection{Effect of inhomogeneity: deviations from thermality\label{effectinh}}

\begin{figure}
\includegraphics{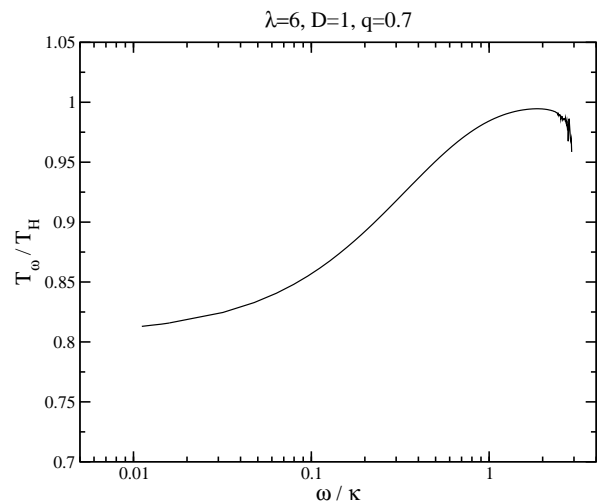}
\caption{Effective temperature as a function of $\om/\kappa$ for $D=1$, $q=0.7$ and $\lambda=6$. This situation describes approximately the experimental realization of Ref.~\cite{Steinhauer:2009aa}. \label{fig::Jeff}}
\end{figure}

In the preceding sections, we fixed $q=0.3$, thus restricting attention to experimental cases where the variation of $c+v$ is mainly due to the sound speed. This would be the case for instance if one used a Feshbach resonance to vary the coupling constant $g$ across the BEC. In~\cite{Steinhauer:2009aa}, the authors report to have created a WH/BH horizon pair in a BEC using a different technique, where a local increase in the flow velocity leads to a decrease in the speed of sound. Their profile $c+v$ is not symmetric with respect to the horizon, contrary to our parameterization, and has two horizons. Ignoring the WH horizon we can approximately describe their experimental realization within our setup. From their experimental values one gets $\lambda\simeq 6$, $D \simeq 1 $, and $q \simeq 0.7$, which yield $\om_{\rm max}/\kappa\simeq 3$. Given the results of the preceding section, one expects a robust spectrum of phonon radiation. This is however not the case, as shown in \figr{fig::Jeff} where the effective temperature $T_\om$ is shown as a function of the frequency. The temperature becomes equal to $T_H$ only for frequencies $\om \simeq \kappa$. In the low frequency part the power is suppressed. From this case we learn that the precise mixture of $c$ and $v$ used to form the horizon significantly affects the properties of the Hawking flux and that the radiation produced in the setup of Ref.~\cite{Steinhauer:2009aa} should not have a thermal spectrum.

\begin{figure}
\includegraphics{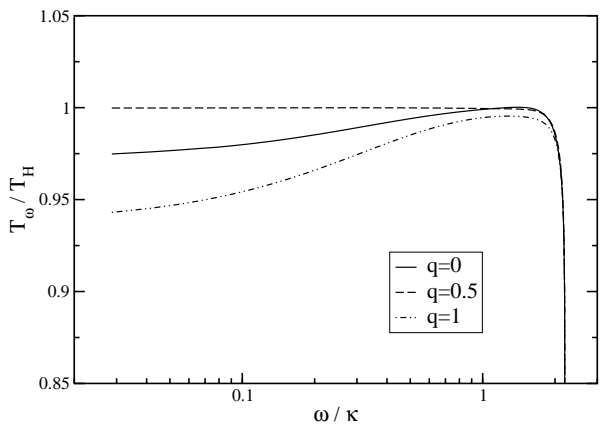}
\caption{Effective temperature as a function of $\om/\kappa$ for $D=0.4$ and $(q,\, \lambda)=(0,\, 19.54)$, $(0.5,\, 17.15)$, $(1,\, 15.47)$. $\om_{\rm max}/\kappa$ is the same for all curves, equal to $2.2$.\label{fig::effetinhf}}
\end{figure}

The effect of $q$ is investigated more systematically in \figr{fig::effetinhf}, for a lower value of $\om_{\rm max}/\kappa$ so as to ensure better numerical control. In that figure, $T_\om/T_H$ is plotted versus $\om/\kappa$ for $D=0.4$, and $q=0$ (homogeneous BEC where only $c$ varies), $q=0.5$ and $q=1$ (only $v$ varies). For each value of $q$, $\lambda$ is tuned so that we work at fixed $\om_{\rm max}/\kappa$ to facilitate the comparison. For $q=0$ and $q=1$, the effective temperature varies significantly and the radiation is thus not thermal. For $q=0.5$, as in the previous plots with $q=0.3$, the radiation is (almost) thermal. We verified that the same results hold with other values of $D$ and $\lambda$, with the same particular role of $q=0.5$.

The strong effect of $q$ on the Hawking flux can be understood as follows. 
Contrary to the setup studied in~\cite{Macher:2009tw} where the $u$ and $v$ sectors were completely decoupled
in the dispersionless limit, 
there is no such decoupling in the present settings. The creation of left-moving quanta and the elastic scattering of $u$ quanta into $v$ quanta will thus be large, even when the dispersion plays no role. 
The above results thus suggest that for $q$ close to 0.5, 
the $u$-$v$ mixing is small and gets larger for extreme values of $q$ near $0$ and $1$. To confirm 
this, we now turn to the properties of the flux of left-moving quanta, $\bar n^v_\om$.

\subsubsection{Mixing of $u$ and $v$ quanta\label{numresnv}}

\begin{figure}
\includegraphics{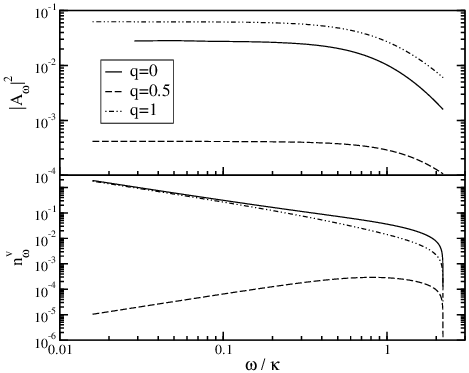}
\caption{$|A_\om|^2$ (upper plot) and $n^v_\om$ versus $\om/\kappa$. The parameters are identical to those of \figr{fig::effetinhf}.\label{fig::effetinhv}}
\end{figure}

Figure~\ref{fig::effetinhv} shows the particle flux of $v$ quanta $\bar n^v_\om$ and the elastic scattering coefficient $|A_\om|^2$ as a function of $\om/\kappa$ for the same parameters as in \figr{fig::effetinhf}. 

At low frequencies, $|A_\om|^2$ is nearly constant. It is two orders of magnitude smaller for $q=0.5$ than for the other two values. 
It should also be noticed that $|A_\om|^2$ does not vanish for $\om \to \om_{\rm max}$ since both $\phi^{u}_\om$ and $\phi^v_\om$ remain well-defined above $\om_{\rm max}$ so that 
$|A_\om|^2$ connects smoothly to $|R_\om|^2$ at $\om=\om_{\rm max}$. 

Near $\om_{\rm max}$, 
$\bar n^v_\om$ goes to zero in the three cases. 
Instead, the low frequency behavior of $\bar n^v_\om$ changes dramatically depending on the value of $q$. 
First, the curves corresponding to $q=0$ and $q=1$ join at low frequencies, 
and are proportional there to $1/\om$. 
This behavior in $1/\om$ means that the energy flux $\hbar\om \, \bar n_\om^v$ 
carried by the $v$ quanta is constant and nonvanishing at low frequencies. 
Secondly, for $q=0.5$, $\bar n^v_\om$ stays everywhere below $10^{-3}$ and is proportional to $\om$ at low frequencies. This proportionality with $\om$ was also obtained in the different setup of Ref.~\cite{Macher:2009tw}, 
and seems to indicate that the case $q=0.5$ (and not $q=1$ as one would have thought naively) is 
effectively  similar to the setup of that reference. 
Even though this particular point is most probably an artefact of \eqr{vdD}  
where $v$ and $c$ follow the same function, the 
qualitative conclusion that the $u$-$v$ mixing is lower when the horizon is formed 
by a variation of both $v$ and $c$ rather than of only one of them, 
should hold experimentally. We emphasize this point because a small $u$-$v$ mixing guarantees that the Hawking $\cal B$ type of 
correlations is the largest, as explained at the end of Sec.~\ref{spatialstruct}. 

\begin{figure}
\includegraphics{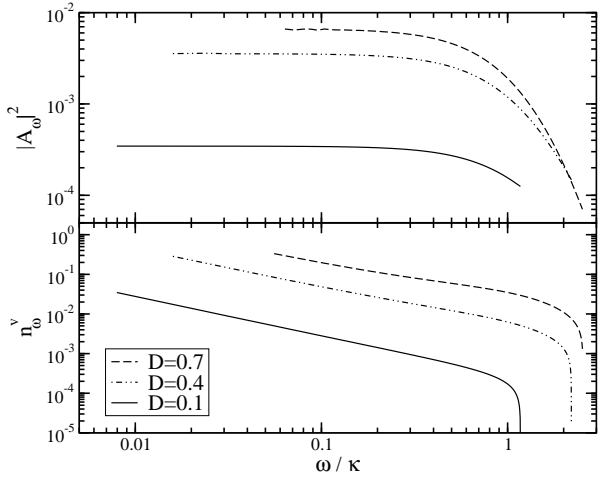}
\caption{ $|A_\om|^2$ (upper plot) and $\bar n^v_\om$ versus $\om/\kappa$, for various values of $D$. The legend applies to both plots. The values of $\lambda$ are identical to those of \figr{fig::typicalfom} and $q$ is fixed to $0.3$.\label{fig::uvmixDeffect}}
\end{figure}

Figure \ref{fig::uvmixDeffect} shows the influence of $D$ on the $u$-$v$ mixing coefficients $\bar n^v_\om$ and $|A_\om|^2$, for $q=0.3$. Both the creation of $v$ quanta and the elastic scattering coefficient $A_\om$ are significantly affected by $D$ and increase by more than one order of magnitude between $D=0.1$ and $D=0.7$. 
This will have important consequences 
when taking into account an initial temperature.

\subsubsection{Dispersionless elastic scattering}

\begin{figure}
\includegraphics{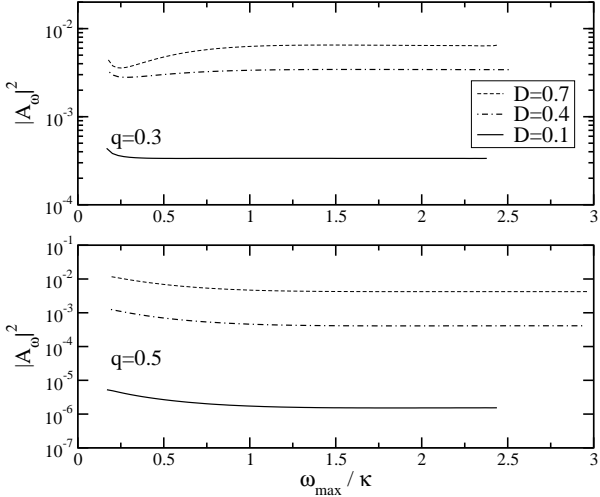}
\caption{$|A_\om|^2$ evaluated for $\hbar\om=k_BT_H$, versus $\om_{\rm max}/\kappa$ for $D=0.1$, $D=0.4$ and $D=0.7$. $q$ is fixed to $0.3$ in the upper plot and $0.5$ in the lower one.\label{fig::robustnessA}}
\end{figure}

It is instructive to look at the behavior of $|A_{\om_H}|^2$ evaluated for $\hbar\om_H=k_BT_H$ as a function of $\om_{\rm max}/\kappa$, and for fixed values of $D$ and $q$. 
It is shown in \figr{fig::robustnessA}. When $\om_{\rm max}/\kappa$ is greater than about $2$, $|A_{\om_H}|^2$ becomes nearly constant, with a value that depends on 
$D$ and $q$. This means that for large $\om_{\rm max}/\kappa$, the dispersion no longer plays 
any role. 
This is to be opposed to what was found in~\cite{Macher:2009tw}, where $|A_{\om_H}|^2$ scaled approximately as $\lambda^{-4}$. The reason is that in a BEC, the wave equation~\eqr{centraleq2} does not factorize into a $u$ and a $v$ part in the dispersionless limit $\lambda\to \infty$. 
Thus, there is always some elastic scattering between $u$ and $v$ modes. What \figr{fig::robustnessA} shows is that the dispersionless value of $|A_{\om_H}|^2$ is quickly reached, or equivalently that the $\lambda$-dependent contributions 
vanish rapidly for increasing $\lambda$. This $\lambda$-independent elastic scattering exists also for $q=0.5$, whereas we have seen that in this case $\bar n^v_\om$ has a behavior similar to the corresponding one in~\cite{Macher:2009tw}. This is somewhat surprising.

Note finally that, since $|A_\om|^2$ is nearly constant in $\om$, 
see \figr{fig::effetinhv} and \figr{fig::uvmixDeffect}, 
the value $|A_{\om_H}|^2$  for some $(D,q)$ 
in the regime where $\lambda$ plays no role actually 
gives $|A_{\om}|^2$ for all $\om$.

\subsubsection{Strongly dispersive regimes\label{dispersive}}

In \figr{fig::robustness}, we saw that, starting in the robust regime $\om_{\rm max}/\kappa>2$ and reducing $\om_{\rm max}/\kappa$, $\Delta_H$ first increases, which indicates also an increasing asymptotic temperature $T_0$, and then monotonically decreases for $\om_{\rm max}/\kappa \lesssim 0.6$. We refer to the latter regime as the strongly dispersive regime. It is reached when the variation of $c+v$ occurs on much smaller distances than what is assumed in \eqr{constrLD}. 

\begin{figure}
\includegraphics{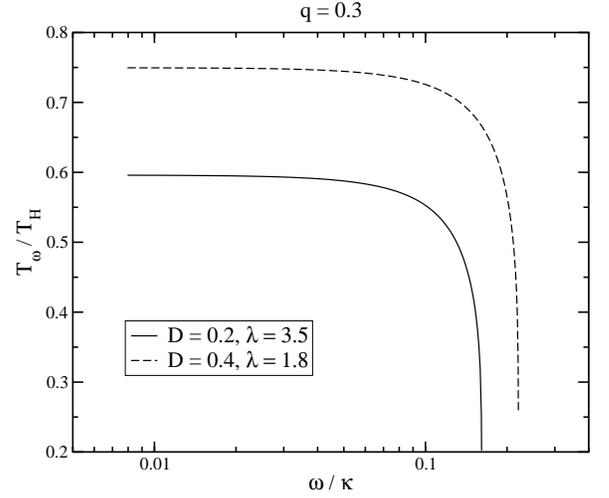}
\caption{Effective temperature $T_{\om}$ divided by Hawking temperature
as a function of $\om/\kappa$ for $q=0.3$ and $(D,\lambda)=(0.2,3.5)$ (solid line) 
and $(0.4,1.8)$ (dashed line)
when the variation of $c+v$ occurs on one healing length.\label{fig::dispersive}}
\end{figure}

In \figr{fig::dispersive}, the variation is assumed to occur on one healing length. This amounts to change the numerical factor in \eqr{constrLD} so that now $\Lambda/\kappa = 0.7/D$. The effective temperature $T_\om$ is shown for $q=0.3$ and $(D,\lambda)=(0.2,3.5)$ and $(0.4,1.8)$. The corresponding values of $\om_{\rm max}/\kappa$ are $0.16$ and $0.22$, that is, much smaller than in \figr{fig::typicalfom}. The shape of the two curves is remarkably similar to what was obtained in the right plot of \figr{fig::typicalfom}: 
 as soon as the frequency is less than about $0.2\times\om_{\rm max}$, $T_\om$ is nearly constant 
(the running $\Delta_{T_0}$ is equal to $7\%$ and $5\%$) and then drops quickly to zero when approaching $\om_{\rm max}$. However, $T_0$, the asymptotic value of $T_\om$, significantly differs
from the Hawking temperature: it is respectively equal to $0.6\, T_H$ and $0.75\,T_H$.
From this we conclude that when dispersion is strong, 
the low-frequency spectrum remains in $1/\om$ as for a Planck spectrum. However since the cutoff frequency $\om_{\rm max}$ is only $\hbar \om_{\rm max}/(k_B T_0) = 1.7$ and $1.9$ respectively, the spectrum is no longer Planckian.~\footnote{We thank T. Jacobson for a remark about this point.} We  finally notice that $T_0/T_H$ becomes smaller and smaller as $\om_{\rm max}/\kappa$ decreases. 

\begin{figure}
\includegraphics{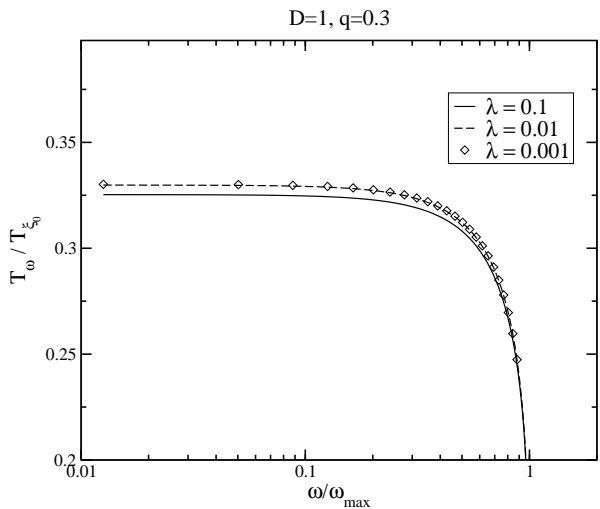}
\caption{Effective temperature $T_{\om}$ divided by the healing temperature of \eqr{Thealing}, 
as a function of $\om/\om_{\rm max}$ for $q=0.7$ and $D=1$ in extremely dispersive cases, when $\kappa \gg c_0 / \xi_0$. 
\label{fig::verydispersive}}
\end{figure}

It is thus interesting to consider extremely dispersive cases where $c_0\kappa^{-1}$ 
is much smaller than $\xi_0$. In \figr{fig::verydispersive} we represented $T_\om/T_{\xi_0}$ as a function of $\om/\om_{\rm max}$, with $D=1$, $q=0.3$ and small values of $\lambda$ from $0.1$ to $10^{-3}$, corresponding to $\om_{\rm \max}/\kappa$ ranging 
from $4\times 10^{-2}$ to $4\times 10^{-4}$. $T_\om$ still tends to a constant value $T_0$ at low frequencies and the low-frequency part of the spectrum still behaves as $\om^{-1}$. 
The ratio $\hbar \om_{\rm max}/(k_B T_0)$ is nearly constant, equal to $1.6$ for all three curves. The running is large, of the order of $12\%$. 
An important result is that the ratio $T_0/T_{\xi_0}$ saturates at a constant value as $\om_{\rm max}$ decreases.~\footnote{This result is corroborated by 
\cite{Recati:2009aa} where a step-like variation (in $x$) of the sound speed is considered in a homogeneous condensate flowing with a constant $v$ so as to get a transition from a sub- to a supersonic flow. This case can be handled analytically and, in our language, it corresponds to the limit $\kappa \to \infty$, with $D,\, \xi_0$ fixed and $q=0$.}
This demonstrates that $\kappa$ becomes irrelevant and that $T_0$ is fixed only by $\xi_0$. 
We have verified that this result also holds for other values 
of $D$ and $q$, but with different asymptotic values for $T_0/T_{\xi_0}$.

\subsection{Spectral properties with an initial temperature\label{temperature}}

\subsubsection{Energy spectrum\label{fomT}}

\begin{figure*}
\includegraphics[scale=0.9]{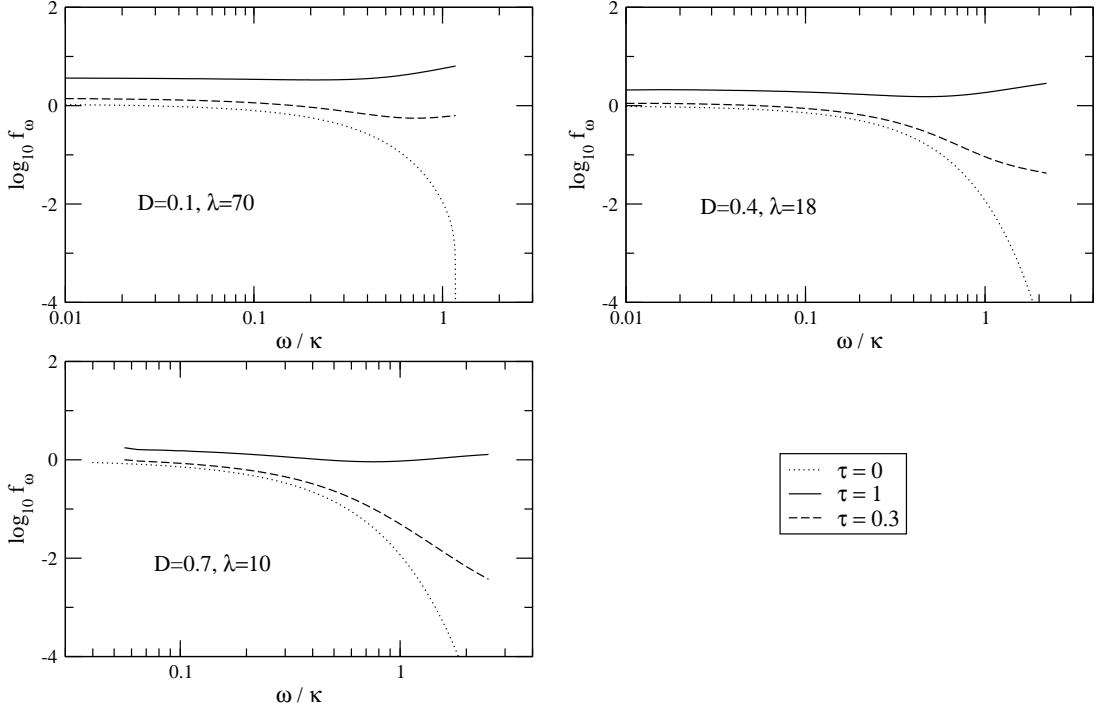}
\caption{Energy flux emitted to the right of the horizon. $q$ is fixed to $0.3$. The dotted spectra are those of \figr{fig::typicalfom}. 
The values $\tau=1$ (solid lines) and $\tau=0.3$ (dashed lines) correspond respectively to an initial temperature $T_{in}=30\,{\rm nK}$ and $T_{in}=10\,{\rm nK}$, see \eqr{tau}.\label{fig::typicalfomWithT}}
\end{figure*}

As pointed out in Secs.~\ref{Tini} and~\ref{numbers}, 
the residual temperature of a condensate is expected to be about two orders of magnitude higher than $T_H$. 
The spectra of the previous sections are thus unlikely to be observed as such 
and it is necessary to include the effects of an initial temperature.  
This is easily done using \eqr{betathreeoccn} 
and the numerical values of the coefficients of the Bogoliubov transformation. 
Assuming that the three initial occupation numbers are characterized by a common comoving temperature, 
 they are given by \eqr{thermal}. 
Their calculation reduces to the computation of the functions $\Om^{in}(\om)$ for the three types of modes, which is easily done by solving \eqr{reldisp}. Since 
$\Om^{in,\, a}$ are three nontrivial functions of $\om$, the initial distributions 
are not Planckian in $\om$.

The energy spectrum emitted to the right of the horizon, with a nonzero initial temperature, is defined as
\be
f^{\rm fin}_\om = 2\pi\frac{\om}{\kappa}n^{\rm fin}_\om.
\ee
It is represented in \figr{fig::typicalfomWithT} for the same set of parameters as in \figr{fig::typicalfom}. Two values of the initial temperature are considered: a conservative one, $T_{in}=30\,{\rm nK}$, and an optimistic one, $T_{in}=10\,{\rm nK}$. With the notations and the constraints of Sec.~\ref{numbers}, they correspond to $\tau=1$ and $\tau=0.3$ respectively. For $D=0.1$, $D=0.4$ and $D=0.7$ the ratio $T_{in}/T_H$ is respectively $300$, $75$ and $43$ when $\tau=1$, and $90$, $23$ and $13$, when $\tau=0.3$. 

The spectra differ greatly from those at zero temperature, and have a nontrivial behavior. Without surprise, they no longer vanish when approaching $\om_{\rm max}$ 
because neither $\bar n^{in}$ nor $|A_\om|^2$ do. 
The most interesting point is that for $D=0.4$ and $0.7$ and the lower value of $\tau$, the spectra follow relatively closely the zero-temperature ones until $\om\simeq\kappa$. Thus, for frequencies below $\kappa$, large values of $D$, and low 
initial temperatures, the measure of the  phonon energy spectrum 
to the right 
gives a good estimate of the zero-temperature flux (i.e., Hawking radiation).

\begin{figure}
\includegraphics{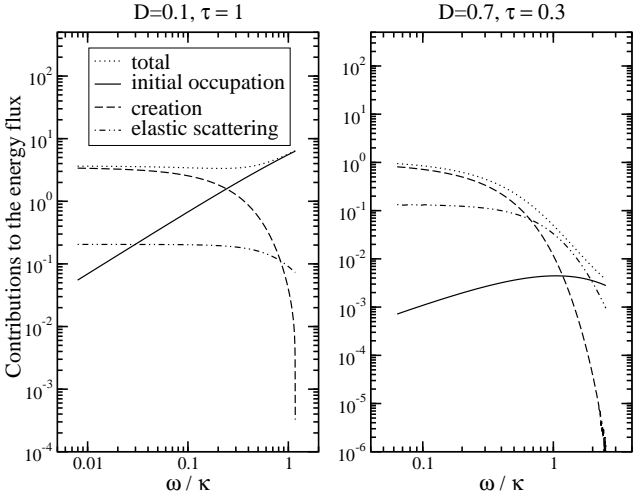}
\caption{Different contributions to the energy flux to the right of the horizon. Left plot: $D=0.1$, $\tau=1$. Right plot: $D=0.7$, $\tau=0.3$.\label{fig::components}}
\end{figure}
To  better understand these spectra, we show in \figr{fig::components} 
the contributions to the energy flux of each of the three terms in \eqr{betathreeoccn}, 
along with the full spectrum,  in the two extreme cases $D=0.1$ with $\tau=1$, and $D=0.7$ with $\tau=0.3$.

In both cases, at high frequencies, the initial distribution $\bar n^{in}_\om$ largely dominates the spectrum, 
and all we see is just the flux of the initial quanta. 
At low frequencies instead, 
the main contribution comes from the 
spontaneous plus stimulated emission term $|\beta_\om|^2 (1+ \bar n^{in}_\om+ \bar n^{in}_{-\om})$, 
called ``creation term'' in \figr{fig::components}. 
For $D=0.1$, the $u-v$ 
scattering term is small and never dominates, but it is never completely negligible. 
The contribution from $|A_\om|^2\bar n^{in, \, v}_\om$  
quickly becomes comparable to the exponentially decreasing 
stimulated emission.

\begin{figure*}
\includegraphics{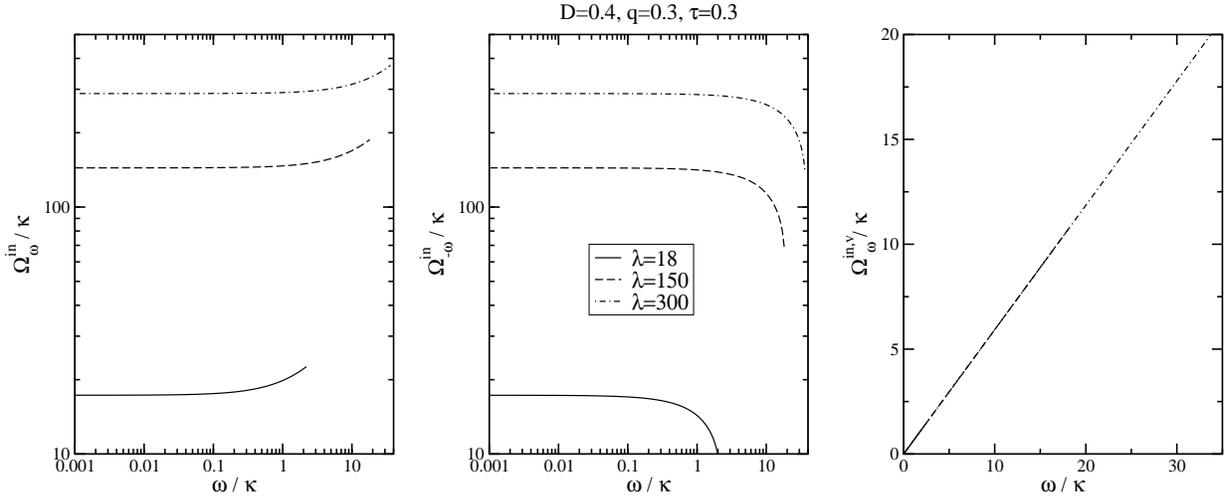}
\caption{Initial proper frequency $\Om^{in}/\kappa$ as a function of $\om/\kappa$ for the $u,\om$ (left plot), $u,-\om$ (middle plot) and $v,\om$ modes (right plot), for $D=0.4$, $q=0.3$ and $\tau=0.3$, and different values of $\lambda$. The legend applies to all plots.\label{fig::Omin}}
\end{figure*}

For $D=0.7$, this contribution is not much smaller than the creation term at low frequencies, as could be expected from  the results of \figr{fig::uvmixDeffect}. 
On the other hand, the contribution from $\bar n^{in}_\om$ long remains very small compared to the other two, until about $\om=\kappa$. Thus the creation+stimulated emission term is actually dominated by the Hawking pair creation effect, since 
$\bar n^{in}_{-\om}$ is close to $\bar n^{in}_\om$. 
This explains the relative similarity between the spectra with $\tau=0.3$ and $\tau=1$ in \figr{fig::typicalfomWithT}, when $D=0.7$. 

In brief, the main lesson from these plots is 
that, in general, there is no clear hierarchy between the various contributions in $\bar n^{\rm fin}_\om$, which makes the interpretation of a measurement highly nontrivial. Nevertheless, let us try to identify what the optimal conditions are for such a detection.

\subsubsection{Optimal experimental conditions\label{numOmin}}

\begin{figure*}
\includegraphics{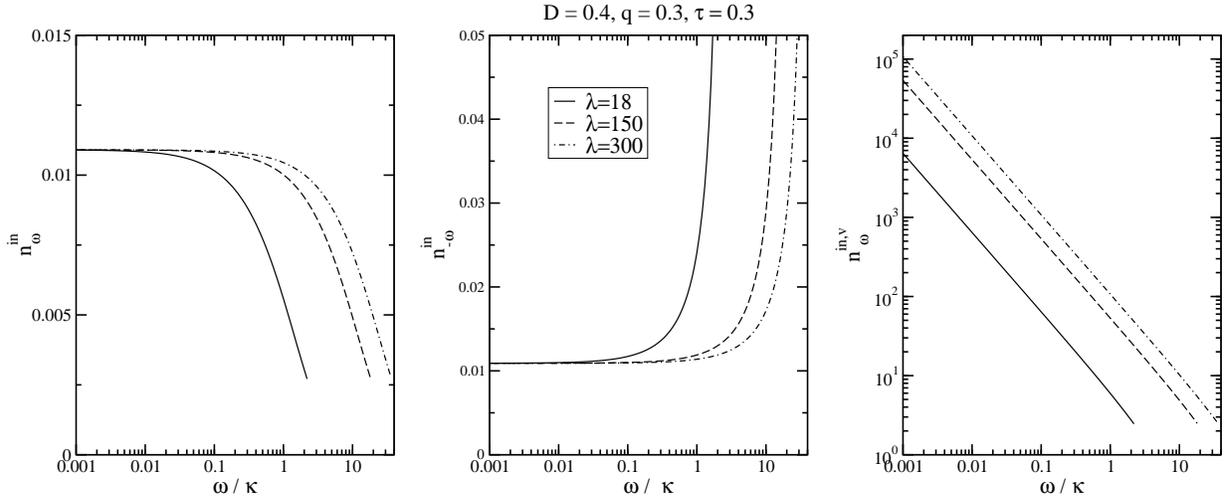}
\caption{Initial occupation numbers $n^{in}_\om$ (left plot), $n^{in}_{-\om}$ (middle plot) and $n^{in,v}_\om$ (right plot) versus $\om/\kappa$, for the same values of the parameters as in \figr{fig::Omin}.\label{fig::nin}}
\end{figure*}

The comoving frequency $\Om$ of the right-moving quanta is strongly redshifted in the black hole geometry, which implies
that the initial $\Om^{u, in}(\om)$ is much larger than $\om$. Since the integrated redshift increases with $\lambda$, 
one could ask whether lowering the Hawking temperature with respect to the initial temperature $T_{in}$, with a fixed value of $D$, could not improve the results above. Indeed, if the increase of the redshift were such that $\Om^{in}/T_{in}$ would grow with $\lambda$, the initial occupation numbers $n^{in}_\om$ and $n^{in}_{-\om}$ would decrease exponentially, and one could hope to have the stimulated plus creation term dominate over the other two on a larger interval of frequency $\om$. This is not the case however, as we now explain.

The initial proper frequency $\Om^{in}(\om)/\kappa$ for the three types of modes 
is shown in \figr{fig::Omin}, for $D=0.4$, $q=0.3$, and different values of $\lambda$. As expected, at a given $\om$, $\Om^{in,u}/\kappa$ is an increasing function of $\lambda$. However it scales only as $\lambda$, and since the ratio $T_{in}/T_H$ also is proportional to $\lambda$, $\Om^{in,u}/T_{in}$ hardly changes. On the other hand, the left movers suffer almost no redshift, and $\Om^{in,v}$ has almost no dependence on $\lambda$. Thus, $\Om^{in,v}/T_{in}$ decreases as $\lambda^{-1}$ and the initial occupation number of the $v$ modes increases as $\lambda$ for a given $\om$. 

These remarks are summarized in \figr{fig::nin}, where the occupation numbers $\bar n^{in}_\om$ 
of \eqr{thermal} 
are represented for the same parameters as in \figr{fig::Omin}. $\bar n^{in}_\om$ and $\bar n^{in}_{-\om}$ 
are almost equal and, more importantly, do not change with $\lambda$ at low frequencies. $\bar n^{in,v}_\om$ on the other hand explodes when $\lambda$ increases. Since $|A_\om|^2$ does not depend on $\lambda$ in first approximation, the consequence 
is that the scattering contribution in $\bar n^{fin}_\om$ 
dominates at low frequencies when $\lambda$ increases, since the other contributions remain unchanged.
In conclusion, there is no gain in lowering $T_H$. 

In fact, as can be seen from the figures of the previous section, the optimal conditions to be able to measure directly the zero temperature flux $f_\om$ are  reached when there is a large relative variation of $c+v$ on the smallest possible distance, that is the largest possible $D$ with the smallest possible $\lambda$. Remember also that $c$ and $v$ should contribute nearly equally to the variation for $|A_\om|^2$ to be as small as possible.

\subsubsection{Density fluctuations and atom flux\label{numresdensfluct}}

\begin{figure}
\includegraphics{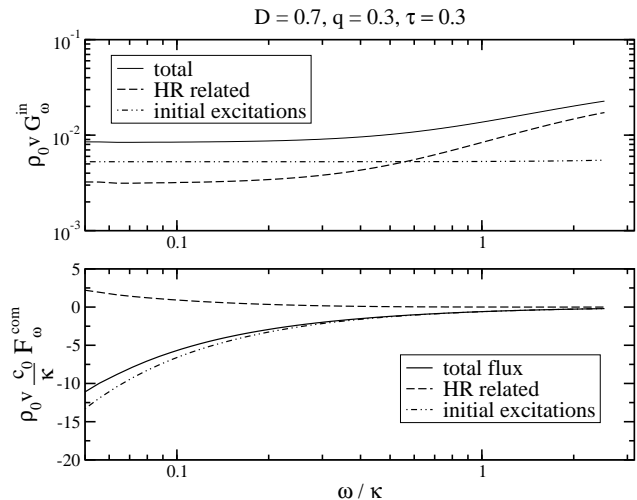}
\caption{Upper plot: adimensionalized (by multiplication by $\rho_0 v$) 
power spectrum of density fluctuations as a function of $\om/\kappa$. 
Lower plot: adimensionalized comoving atom flux as a function of $\om/\kappa$. In both plots, $D=0.7$, $q=0.3$ and $\tau=0.3$, and $G^{\rm in}_\om$ and ${\cal F}^{\rm com}_\om$ are evaluated in the right asymptotic region, in the coincidence point limit.  ``HR related'' refers to the contribution weighted by $n^{fin}_\om$
 and ``initial excitations'' to the one weighted by $n^{v,in}_\om$. 
\label{fig::densfluct}}
\end{figure}

As explained in Sec.~\ref{discussion}, the flux $f_\om$ studied in the previous sections is observable only if one can distinguish experimentally between left- and right-moving phonons. It is thus important to address the question whether the easily accessible density fluctuations also contain some signature of the particle creation process. The function $\rho_0 v G^{in}_\om$ of \eqr{n1}, evaluated in the coincidence point limit in the right asymptotic region, is represented in the upper plot of \figr{fig::densfluct} for $D=0.7$, $q=0.3$ and $\tau=0.3$. This quantity is adimensional,  does not depend explicitly on $\rho_0$ and $c_0$ and is proportional to the power spectrum of the density fluctuations. 
In the lower plot, $\cal F_\om^{\rm com}$, defined in \eqr{flusso} and adimensionalized by multiplication with $\rho_0 v c_0/\kappa$, is represented, for the same set of parameters. In both cases, the contributions governed respectively by $n^{\rm fin}_\om$ (referred to as ``HR related'' in the figure) and by $n^{in,v}_\om$ have been plotted separately. They are of the same order of magnitude but for $\om\ll \kappa$, that is, in the region where $n^{\rm fin}_\om$ is mainly due to Hawking radiation, see \figr{fig::typicalfomWithT}, the contribution from the initial excitations is greater by a factor of about 2 in the density fluctuations, and of up to 6 in the atom flux. This confirms the remark made in the theoretical analysis that no clear signature of Hawking radiation can be observed in these local observables. However, as pointed out in Sec.~\ref{densfluctth}, one could hope to combine both observables to extract $n^{\rm fin}_\om$.

\section{Conclusions}

In this work, we presented a complete description of the scattering of the phonon modes
propagating in stationary, one dimensional condensates that possess a sonic horizon. 
Our description is based on the BdG equation, 
and the scattering in the horizon region is expressed 
in terms of $3 \times 3$ Bogoliubov transformations relating asymptotic one phonon modes. 

We have shown that this scattering affects 
two types of observables, local ones such as the density fluctuations,
and the long distance correlations. The former are governed by expectation values which are diagonal in the occupation number, see \eqr{n1}, whereas the latter are determined by interference terms, \eqr{ABC2}, which reveal the entangled nature of the final state.

When taking into account the condensate temperature, the local observables are in general dominated by the initial distributions, whereas nonlocal  correlations are amplified by these initial distributions without having their pattern modified (when the initial state is uncorrelated). 
Therefore the latter probably constitute the clearest indication that the Hawking effect is taking place. 
However, it is worth noting that the spatial structure of the pattern is not specific to HR strictly speaking, i.e. to a nearly Planckian spectrum, 
since it is determined before all by the structure of the $3\times 3$ transformation of \eqr{Bog} rather than by the value of its coefficients. In fact, similar patterns are found when considering the scattering of classical waves, as explained in App.~\ref{coher}, and in the limiting cases $\kappa\to \infty$ and $\kappa = 0$ (with nonvanishing higher derivatives of $c+v$ at $x=0$ for the latter) where the spectrum is no longer approximately Planckian (as we verified). 

In the last part, we numerically integrated the BdG equation, and obtained the occupation numbers
of the three kinds of phonons, both with and without an initial temperature. 
The main results are the following. 

Firstly, in the initial vacuum and when the $u$-$v$ mixing is small ($q\simeq 0.5$), the spectrum of the outgoing (right-moving) phonons is Planckian, with a temperature determined by the gradient $\kappa$ of \eqr{kappadef}, as soon as $\om_{\rm max}/\kappa\gtrsim 2$ and for frequencies $\om<\om_{\rm max}$, see \figr{fig::typicalfom}. This result, derived directly from the BdG equation (without further approximation), makes precise the domain of validity of the analogy between relativistic fields in black hole metrics and a phonon field in the corresponding ``dumb hole''. When $\om_{\rm max}/\kappa < 2$, the analogy fails even for the lowest frequencies, see \figr{fig::dispersive}, which invalidates the condition usually found in the literature that the analogy is applicable for wavelengths greater than the healing length. When the $u$-$v$ mixing is not negligible, the spectrum deviates from the Planck spectrum, see Figs.~\ref{fig::effetinhf} and \ref{fig::effetinhv}. This is due to grey body factors (that can also be computed to a good approximation using the gravitational  analogy~\cite{Anderson:2009va}).

Secondly, as for the detection of the analog HR in a BEC, if one can distinguish left- from right-moving phonons, the Hawking flux could be observable 
when the relative variation of $c+v$ is large and happens on a small number of healing lengths. Instead, if they cannot be distinguished, local observables are completely dominated by the initial temperature of the condensate. One must then resort to nonlocal observables such as density-density correlations~\cite{Balbinot:2007de,Carusotto:2008ep}. 

Thirdly, in App.~\ref{appWH}, we established the relationship between the fluxes emitted by
white and black hole flows. Because of the blueshift effect in WH flows, the correlation pattern is more amplified by the presence of initial quanta than in BH flows and should therefore be easier to detect. This could well be the optimal case for observing HR in the laboratory through the correlation pattern.

\begin{acknowledgments}
R.P. is grateful to Roberto Balbinot for many stimulating discussions over the last years,
as well as his comments on an early version of this paper.
We both would like to thank the participants 
of the workshop ``Towards the observation of Hawking radiation in condensed matter systems'' 
(http://www.uv.es/workshopEHR/) held at IFIC in Valencia in February 2009
for interesting remarks following the presentation of results contained in this work. We are also grateful to Jeff Steinhauer for comments about his work~\cite{Steinhauer:2009aa} and BEC in general, and to Iacopo Carusotto for comments on App.~\ref{coher}.
\end{acknowledgments}

\appendix

\section{Nonstationary case\label{nonstatio}}

In this appendix we consider nonstationary condensates. As we shall see, little is modified when compared with the stationary case studied in the body of the paper. To ease the comparison we still work with the one dimensional case. The extention to the three dimensional case is trivial.
 
In the mean field approximation, the condensate is now described by 
\be
\Psi_0(t,x) = \sqrt{ \rho_0(x, t)}\, 
e^{i W_0(t, x)}. 
\label{eq1td}
\ee
This wave satifies \eqr{GP} where $V$, and $g$ depend both on $x$ and $t$. (As pointed out in \cite{Carusotto:2008ep}, some tuning of $V$ and $g$ might be necessary not to produce time-dependent effects that might hide the Hawking radiation.) The conservation of the number of atoms follows from \eqr{GP}, and gives the continuity equation 
\be
\partial_t \rho_0  + \partial_x( \rho_0  v ) = 0,
\label{conttd}
\ee
where $v(t, x) = \hbar k_0(t, x)/m$ is the velocity of the condensate, and $k_0(t, x) = \partial_x W_0(t, x)$  its wave vector. 
Plugging \eqr{eq1td} into 
\eqr{GP} and using \eqr{conttd} still gives \eqr{tuned} 
where the chemical potential is replaced by $\hbar \om_0(t,x)= - \hbar \partial_t W_0$. Then, because of \eqr{conttd}, the condensate is still characterized by $v(t,x)$ and the speed of sound
\be
c^2(t, x) = g(t, x) \, \frac{\rho_0(t, x)}{m }.
\label{cdeftd}
\ee

To describe the phonon modes, as in Sec.~\ref{settings}, it is convenient to work with the relative fluctuation $\phi$ defined in \eqr{resc}. 
 Then, using Eqs.~(\ref{eq1td}, \ref{cdeftd}), one still gets \eqr{BdG2}. To get the $c$-number modes which shall be used to proceed to the second 
quantization, we look for a complete orthonormal family of doublets $W_j= (\phi_j, \varphi_j)$, where the index $j$ now does not correspond to a conserved quantum number. The scalar product with respect to which orthonormality is defined is still given by \eqr{BECnorm}, with $\rho_0(x)$ replaced by $\rho_0(x,t_0)$. (The choice of the time $t_0$ when the product is evaluated does not change its value since it is conserved.) This family enters $\phi$ as in \eqr{statio}:
\be\label{statiotd}
 \phi(t, x) 
=  \sum_j\left[ \hat a_j \,
\phi_j (t, x)  + \hat a_j^\dagger  \left(\varphi_j(t, x)\right)^*\right],  
\ee
where $\hat a_j$ and $\hat a_j^\dagger$ are annihilation and destruction
operators, 
with which one can construct the phonon Fock space. However, contrary to the case studied in the text, there is in general no clear interpretation of the vectors in this Fock space in terms of particle content, and the notion of phonons is inherently ambiguous. The notion of phonon can be recovered if the condensate is stationary in the asymptotic past and the asymptotic future. In that case, as in the text, one can define two Fock spaces, the in one and the out one.

Inserting \eqr{statiotd} in \eqr{BdG2} and 
taking the commutator with $\hat a_j$ and $\hat a_j^\dagger$ yields:
\be
\begin{split}
\left[ i\hbar \left(\partial_t + v \partial_x \right) -  T_\rho
- m  c^2  \right]\, \phi_j &=  m  c^2 \,  \varphi_j, \\
\left[ - i \hbar \left(\partial_t + i v_{}  \partial_x \right) - T_\rho
- m  c^2  \right]\, \varphi_j &=  m  c^2 \,  \phi_j .
\end{split}
\label{centraleq0td}
\ee
As in the body of the paper, one can eliminate $\varphi_j$ and obtain 
\be
\begin{split}
\Big\{\left[  \hbar \left(\partial_t + v \partial_x \right) -i T_\rho\right]\frac{1}{c^{2}} \left[ \hbar \left(\partial_t + v \partial_x \right)+ i T_\rho\right] \\
   + 2m T_\rho \Big\} \phi_j = 0.
\end{split}
\label{centraleq2td}
\ee
We notice that all kinetic terms are of the form of \eqr{Trho}. 
This is related to the fact that  $T_\rho$ is self-adjoint 
when using the scalar product of \eqr{BECnorm}.

\section{Additional remarks concerning \eqr{centraleq2}\label{additional}}

\subsection{Eikonal approximation and importance of ordering}

It is worth exploring the 
properties of \eqr{centraleq2}, the stationary version of \eqr{centraleq2td}. We first notice that 
\be
\begin{split}
c^2 \left[
 \hbar \left( \om + i v_{} \, \partial_x \right) + T_\rho
\right]
\frac{1}{c^{2}} =
  \left[\hbar \left( \om + i v_{} \, \partial_x \right) + T_\rho
\right] \\
+\, c^2  
\left[ \left(
i\hbar v_{} \, \partial_x  + T_\rho \right)
, \frac{1}{c^{2}} \right]\, .\end{split}
\label{comTc}
\ee
This makes explicit that the spatial gradient 
of $c^2$ 
affects \eqr{centraleq2}
only through a commutator. We also notice that $\varphi_\om$, the other mode of \eqr{statio} obeys 
\be
\begin{split}
\Big\{\left[ -\hbar \left(\om + i v_{} \, \partial_x \right) + T_\rho \right]
\frac{1}{c^{2}} \left[ \hbar \left(\om + i v_{} \, \partial_x \right) + T_\rho
\right] \\
-\, \hbar^2 { v_{} }{ \partial_x \frac{1}{v_{}} }\partial_x
\Big\} \varphi_\om = 0.
\end{split}
\label{centraleq2var}
\ee
The only differences between the equations for $\phi_\om$ and 
for $\varphi_\om$ come from the sign of the commutator between $v_{} \, \partial_x$
and $c^{-2}$. 
When this commutator is neglected, as it is in the WKB approximation, $\phi_\om$ and $\varphi_\om$ thus obey the same equation.

When working to leading order in a WKB (or eikonal) approximation, 
i.e., inserting $\phi_\om \sim e^{i \int \! dx\,  k_\om}$ in \eqr{centraleq2}
(or inserting $\varphi_\om \sim e^{i \int \! dx\,  k_\om}$ in \eqr{centraleq2var})
gives the dispersion relation in a moving fluid of velocity $v_{}$, 
\be
\left( \om -  k v_{} \right)^2 = \Om^2 = k^2 c^2 + \frac{\hbar^2 k^4}{4 m^2} = c^2 k^2 \left( 1 + \frac{\xi^2 k^2}{2} \right) ,
\label{quarticdr}
\ee
where $\xi = \hbar / \sqrt{2}m c $ is the healing length. 
We recover the quartic dispersion relation between the frequency in the comoving frame $\Om = (\om - k v_{} )$, and the wave vector $k$. 
It should be noticed that in a stationary nonhomogeneous flow, $\Om$ and $k$  depend on $x$ 
through $v(x)$ and $c(x)$, whereas the frequency $\om$ is a globally defined constant. 
Remember also 
that $\om$ is not necessarily the lab frequency, because it is defined in the frame in which the condensate quantities only depend on $x$.

Returning to \eqr{centraleq2}, it is clear that its role 
is to fix the exact properties of the ODE obeyed by $\phi_\om$. These properties 
could not have been inferred starting from \eqr{quarticdr}, and applying the
substitution $k \to -i \partial_x$, because 
this naive rule could neither predict the 
ordering of $T$, $v(x)$ and $c^2(x)$ found in \eqr{centraleq2}, nor the different one found 
 in \eqr{centraleq2var}. 

\subsection{Hydrodynamical limit and Euler mode equation}

In the hydrodynamical (dispersionless) limit,  
for long wavelengths with respect to the healing length $\xi$, one can drop the two operators $T_\rho$ in \eqr{centraleq2}. 
Then \eqr{centraleq2} reduces to the Eulerian mode equation which describes sound waves in a moving fluid, and this
with the correct nontrivial ordering of $T$, $v(x)$ and $c^2(x)$. Let us verify this.

The Eulerian action is usually written in terms of the velocity potential $\psi$, 
related to the velocity fluctuation by $\delta v = \partial_x \psi$, see e.g. 
\cite{Balbinot:2006ua}.
To make contact between $\psi$ and 
the 
$\phi$ field, one should compare the fluctuations of $\Psi$ described as in \eqr{resc} to those written as 
\be
\delta \Psi = \delta \left(\rho^{1/2} e^{i W} \right) =
\Psi_0 \left( \frac{\delta \rho}{2\rho_0} + i \delta W
\right) = \Psi_0 \, \phi. 
\ee
Using $\delta v = \frac{\hbar}{m} \partial_x \delta W =  \partial_x \psi$, one obtains
\ba
\phi + \phi^\dagger &=& \frac{\delta \rho}{\rho_0}, \quad \nonumber \\
\frac{\phi - \phi^\dagger}{2 i} &=& \theta =  \delta W  =  \frac{m}{\hbar} \, \psi. 
\label{ReIm}
\ea

Using the phase fluctuation $\theta = m \psi /\hbar $ rather than $\psi$ itself, 
the Euler action is
\be
S_E =  \frac{\hbar^2}{2m}\int dt dx  \rho_0 \left\{ \frac{1}{c^2}\left[\left( \partial_t + v_{} \partial_x\right) \theta \right]^2
- \left(\partial_x \theta \right)^2 \right\} .
\label{Eul1}
\ee
Then, using the continuity equation, the mode equation reads 
\be
\left[ \left(  \partial_t +  v_{} \partial_x\right) 
\frac{1}{c^2}\left( \partial_t  +  v_{} \partial_x\right) 
 - \frac{1}{\rho_{0}}  \partial_x  \, \rho_{0} \, \partial_x  
\right]
\theta = 0 \, .
\label{Eul2}
\ee
When working at fixed $\om= i\partial_t$, and using  $v_{} \rho_0  = {\rm cst.}$,
as announced, one recovers
the dispersionless limit of \eqr{centraleq2} obtained by taking the limit $T_\rho \to 0$.

Notice that \eqr{Eul2} is a generalization of the (dispersionless limit of the) 
mode equation which has been generally 
studied in the literature, see 
Refs.~\cite{Unruh:1994je,Brout:1995wp,Corley:1996ar,Balbinot:2006ua,Macher:2009tw}. 
In those references, the equation also followed from \eqr{Eul1}, 
but with the extra hypothesis that both $c$ and $\rho_0$ can be approximated by constants, 
in which case one recovers the massless relativistic 2D mode equation.

\subsection{Link with Unruh's dispersive models}

It is worth noticing that the dispersive properties of the phonons 
come through the two operators $T_\rho$ in \eqr{centraleq2}. This is not 
what we would have obtained 
had we used the rules of \cite{Unruh:1994je} with a quartic superluminal dispersion. 
Indeed, the Eulerian action 
supplemented by a quartic term ($k^4/\Lambda^2$) is
\be
\begin{split}
S_\Lambda = \frac{\hbar^2}{2m} \int dt dx \rho_0
\Big\{ 
\frac{1}{c^2}\left[\left( \partial_t + v_{} \partial_x\right) \theta \right]^2\\
-\, \left(\partial_x \theta\right)^2 + \frac{1}{\Lambda^2}\left(\partial^2_x \theta \right)^2
\Big\},
\end{split}
\label{Eul3}
\ee
and the corresponding mode equation reads, in a stationary condensate and at fixed $\omega$, 
\be
\begin{split}
\Big[  \left( \om + i v_{} \partial_x\right) \frac{1}{c^2}\left( \om + i v_{} \partial_x\right) - v_{} \, \partial_x \frac{1}{v_{}}  \, \partial_x  \\
+\, \frac{v_{}}{\Lambda^2} \partial^2_x \frac{1}{v_{}}  \, \partial^2_x
\Big]
 \theta_\om = 0.
\end{split}
\label{Eul4}
\ee
In nonhomogeneous situations, the quartic term encoding the dispersion differs from that of \eqr{centraleq2},
which means that $\theta_\om(x)$ will differ from $\phi_\om(x)$ in any nontrivial background.  

Finally, it is also worth noticing that \eqr{centraleq2}
\emph{can} be obtained 
from an action for a single field with a quadratic kinetic term, and which 
generalizes the Euler action: 
\begin{align}
&S_\phi = \nonumber\\ 
&- \frac{\hbar^2}{2m} \int dt dx \rho_0
\Bigg\{ 
\frac{1}{c^2}\left[\left( \partial_t + v_{} \partial_x + 
i \frac{T_\rho}{\hbar}
\right) \phi  \right]^2 
-\, \left(\partial_x \phi\right)^2 
\Bigg\}.
\end{align}
This action can be obtained from $S = \int dt\, d^3x (i \hbar \Psi^* \partial_t \Psi - H)$, where $H$ is given in \eqr{Hsc}, using \eqr{resc}, keeping all quadratic terms in $\phi, \phi^*$, 
and using \eqr{BdG2} to eliminate $\phi^*$ in favor of $\phi$. The conjugate momentum that enters the equal time commutator $[\phi(x), \pi(y)] = i \hbar \delta(x-y)$ is rather unusual 
\be
\begin{split}
\pi &= - \frac{\hbar^2 }{m} \frac{\rho_0}{c^{2}} \left(\partial_t + v_{} \partial_x + i \frac{T_\rho}{\hbar} \right) \phi \\
&= i\hbar  \rho_0 \left( \phi + \phi^\dagger \right) = i\hbar \, \delta \rho,
\end{split}
\ee
where we used \eqr{ReIm}. 
It thus obeys $\pi = -\pi^\dagger$. Using \eqr{statiotd}, $\pi$ decomposes as \be\label{statiotd2}
 \pi(t, x) 
=  \sum_j \left[ \hat a_j \, \pi_j (t, x)  + \hat a_j^\dagger  \left(  \bar \pi_j(t, x)\right)^* \right],  
\ee
where $\pi_j = i\hbar \rho_0 (\phi_j + \varphi_j) = - \bar \pi_j$. The conserved scalar product (which generalizes the standard Klein-Gordon one~\cite{Wald:1995yp} and which agrees with \eqr{BECnorm}) is 
\be
(\phi_2 \vert \phi_1) = \frac{i}{\hbar} \int dx \left( \varphi_2^* \, \pi_1 - \bar \pi_2^*\, \phi_1 \right). 
\ee
It is not clear whether this alternative way of describing the phonon field presents any advantage over the original version of \eqr{centraleq0}. It could nevertheless
be useful to study how dispersion affects the analogy 
with gravitational systems.~\cite{Balbinot:2006qe,Jain:2007gg,Weinfurtner:2008if}

\section{Correlation patterns in the classical limit\label{coher}}

Rather than using the vacuum or a thermal state as in the text, we assume that the initial state also contains a highly excited coherent 
state.\footnote{We are grateful to W. Unruh for bringing our attention to coherent states while one of us, 
R.P., was presenting this work at the Peyresq conference of Cosmology in June 2009.} This state can be 
obtained by making use of the displacement operator  
\be
\hat D_w = \exp\left({w\,  \hat a^{\dagger}_\om - w^* \,  \hat a_\om}\right).
\label{Dw}\ee
Any of the three initial operators $a^{in, j}_\om$ appearing in \eqr{inpowers} can be used
to construct the corresponding initial wave. One can also consider wave packets 
engendered by 
\be
\hat  D_{d, \, w} = \exp{\left[ 
\int \! d\om \left( w\, d_\om \, \hat a^{ \dagger}_\om - w^*\, d_\om^*  \, \hat a_\om\right)\right]}.
\ee
When imposing the normalization $\int d\om \vert d_\om \vert^2 = 1$, the mean occupation number
of initial quanta added by $\hat  D_{d, \, w}$ is $\vert w \vert^2$, as it is for $\hat  D_w$
in \eqr{Dw}.

When $\hat \rho^{\rm in}$, the initial state without the coherent state, is such that 
Tr$[\hat \rho^{\rm in} \, \hat \chi(t,x)]= 0$, the anticommutator 
in the presence of the coherent state separates as
\ba
G(t,x;t', x') &= & \frac{1}{2} {\rm Tr}\left[\left( \hat D_{d, \, w}\,  
\hat \rho^{\rm in} \hat D_{d, \, w}^\dagger \right)
 \left\{\hat \chi(t,x), \,  \hat \chi(t',x')\right\}  \right],
\nonumber \\
&=& \bar \chi_d(t,x) \, \bar \chi_d(t,x) + G^{\rm in}(t,x;t', x'), 
\label{decomp}
\ea
where $G^{\rm in}$ is given in \eqr{Gac}, and where the mean value of the field operator is 
\ba
\bar \chi_d(t,x) &=& {\rm Tr}\left[ \left(  \hat D_{d, \, w}\,  \hat \rho^{\rm in}
 \hat D_{d, \, w}^\dagger \right) \hat \chi(t,x) \right] \nonumber \\
&=&
 \langle 0_{in} | \hat D_{d, \, w}^\dagger  \,\hat \chi(t,x)\, \hat D_{d, \, w} | 0_{in} \rangle ,
\ea
as if the initial state was the in vacuum. 
(To get these equations, we have used Tr$[\hat \rho^{\rm in} \, \hat \chi(t,x)]= 0$ and the relation
$\hat D_{d, \, w}^\dagger  \, \hat a_\om \hat D_{d, \, w} = \hat a_\om  + w d_\om$, see e.g. \cite{Leonhardt:book}.)
From the decomposition \eqr{decomp}, 
we see that the correlation pattern is the sum of the pattern encoded in the state $\hat \rho^{\rm in} $
we studied in the former sections, plus the new pattern encoded in 
the real wave packet $\bar \chi_d(t,x)$ associated with the coherent state.

Assuming that this wave packet is initially made only with $\hat a^{in, \, u}_\om$,
it is given by
\be
\bar \chi_d(t,x) = \int_0^\infty d\om \left[ w d_\om \, e^{- i \om t} \chi^{in, \, u}_\om(x) + c.c. \right].
\ee
At early times, before it enters the near horizon region, 
it describes a single wave packet propagating against the flow in the region where the flow is supersonic,
see \figr{fig::uinom}. As shown in that figure, at late time, it splits into three wave-packets, 
\be
\begin{split}
\bar \chi_d(t,x) = \int_0^\infty d\om \left[ w d_\om 
e^{- i \om t} \, \alpha_\om  \chi^{out, \, u}_\om(x) + c.c. \right]
\\
 + \int_0^\infty d\om \left[ w d_\om   e^{- i \om t}\,  \tilde A_\om \chi^{out, \, v}_\om(x) + c.c. \right]
\\
 + \int_0^{\om_{\rm max}} d\om \left\{ w d_\om   e^{- i \om t}\, 
\beta_{-\om} \left[\chi^{out, \, u}_{-\om}(x)\right]^* + c.c. \right\}.
\end{split}
\label{3ltwp}
\ee
The first one describes the transmitted wave. 
It is amplified with respect to the initial wave
if $\int d\om \vert  \alpha_\om d_\om\vert^2 > 1$ which needs not be always the case, as can be seen from \eqr{HK}.
The second wave describes the left moving packet obtained by elastic scattering. 
The third wave is due to stimulated pair creation process, see the minus sign in  \eqr{HK}. It describes the partner's wave, and  is present only if $d_\om$ has non-vanishing components for $\om < \om_{\rm max}$ since $ \beta_{-\om} = 0$ for $\om > \om_{\rm max}$.
Notice that the Fourier component multiplying the mode $e^{i \om t} \chi^{out, \, u}_{-\om}$ 
is the complex conjugate of $ \beta_{-\om} w d_\om $.  The fact that $d_\om^*$ appears guarantees that when replacing $d_\om $ by $d_\om e^{i \om t_0}$, the partner wave with negative $\om$ stays synchronized, i.e. it is shifted by the same lapse of time as the two other wave packets with positive $\om$. Notice also that the presence of the coefficient $\beta_{-\om}$ implies that the partner wave function cannot be localized in a region smaller than $c/\kappa$ because $\beta_{-\om} \sim e^{- \pi \om /\kappa}$ for $\om \gg \kappa$. In this we recover the width of the correlations to the partner which is found when using the correlation function~\cite{Carlitz:1987aa,Balbinot:2007de}. 

We can now relate the space-time pattern encoded in $G^{\rm in}$, the second term in \eqr{decomp}, to that encoded in $\bar \chi_d$. 
Since these two terms have a completely different origin, one might {\it a priori} think that the two 
patterns will be radically different. However, this is not the case. 

Two conditions must be met for the patterns to be approximately the same. The first concerns the out modes, whereas the second concerns the Bogoliubov coefficients appearing in \eqr{3ltwp}. At a given late time $t$, i.e. after a lapse of time $\Delta t$,  
the chosen wave-packet $\bar \chi_d$ has been scattered, the pattern encoded in $\bar \chi_d$ is given by the three values of $x$ where the waves constructively interfere. Working far from the horizon, i.e. $\kappa \Delta t \gg 1$ (see also the discussion in the paragraph after \eqr{phA}), and using \eqr{HJta},  these are the solutions of
\be
\Delta t^{HJ, \, a}_{\om_d}(x) = \Delta t ,
\label{HTJa2}
\ee
where $\om_d = \int d\om \om \vert d_\om\vert^2$ 
is the mean frequency of the wave packet. When the dispersive effects are weak at late time, as it is the case for $\om_d \ll c/ \xi_0$, 
see the discussion after \eqr{vgr2}, the solutions of \eqr{HTJa2} are essentially independent of $\om_d$. 
Therefore, in the large $x$ limit, one finds the {\it same} pattern of correlations
whether one considers the two-point function at some given time $t$ as a function of $x'$ given $x$, see \eqr{3HJt}, or whether one looks for the two partner waves 
when using a packet which contains a branch that arrives at $x$ at time $t$. 

In the above reasoning, we have neglected the $\om$ dependence of the Bogoliubov coefficients. If their relative phase would vary rapidly, and therefore that of 
$z_\om = \beta_{-\om}/\alpha_\om$ as well, \eqr{HTJa2} could receive
significant corrections that might depend on the value of $\om_d$,
thereby giving different patterns for different wave packets.
This is not the case. When using a massless two dimensional relativistic field, 
one finds that the pattern is universal as $\arg z_\om$ is truly independent of $\om$, 
see e.g. Eq.~(3.49) in \cite{Brout:1995rd}. Upon considering dispersive fields, 
when both $\om_d \ll c/ \xi_0$ and $\kappa \ll c/\xi_0$ 
are satisfied, one finds~\cite{Brout:1995wp} that the late time pattern is unmodified
even though the early pattern is completely different
and depends on both the value of $\Lambda = c/\xi_0$ and the sub- or superluminal character of dispersion~\cite{Balbinot:2006ua,Jacobson:2007jx}. 
These results were obtained using saddle point and WKB approximations which are both reliable in the present regime. 
The analysis was performed with a scalar dispersive field, but it also applies to a phonon field in a BEC because 
the differences between the various dispersive models, see Appendix \ref{additional}, are not relevant in a WKB regime.

In brief, we saw that the late time pattern obtained from the scattering of classical waves is (to leading order in $\om/\Lambda$) the same as that obtained from amplifying vacuum (or thermal) fluctuations. This correspondence follows from the fact that both patterns are obtained from the (common) decomposition of in modes into out modes.  Given this we can address a question which has been raised by several of our colleagues: 
to what extent observing the scattering of classical waves could be considered as a signature of the Hawking effect?

A convincing signature \cite{Rousseau:2008aa} would consist in measuring the contribution of the negative frequency modes which anomalously contribute to \eqr{HK}, once sent a wave-packet composed only of positive $\om$, as it is the case in \eqr{3ltwp}. The experimental difficulty is different for BH and WH flows. In BH flows, one has to be sure that the initial packet contains no negative modes otherwise their final contribution will be dominated by the initial one, thereby preventing a neat measure of $\beta_{-\om}$.  The values of $\om$ with $\beta_{-\om} \neq 0 $ are such that $|\om|<\om_{\rm max}$. Given \eqr{reldisp}, they correspond to an interval of initial values for $k(\om)$:
\be
k_{\rm max}= k(-\om_{\rm max}) < k(\om) < \bar k_{\rm max}= k(\om_{\rm max}),
\ee
where
\be
\bar k_{\rm max} = k_{\rm max} + \sqrt{k_{\rm max}^2+\frac{4m^2}{\hbar^2}\frac{\om_{\rm max}^2}{k_{\rm max}^2}}.
\ee 
The restriction to modes with positive frequency 
imposes $k > k_0 = k(0)= \frac{2m}{\hbar} ({v_-^2-c_-^2})^{1/2}$. By a numerical analysis, we found that the ratio $(\bar k_{\rm max}- k_0)/ k_0$  is of the order of $20\%$, independently of $m$ (or $\xi$), and weakly varying with $D$ and $q$. Thus in principle, one could selectively excite  (e.g. by Bragg spectroscopy techniques~\cite{Steinhauer:2003aa,Ozeri:2005aa})
certain values of $k$ in that interval. However in experiments, there will be a tension between the limited lifetime of the condensate and the narrowness of the 
packets in $k$ space which gives broad packets in real space with long traveling times. Orders of magnitude taken from \cite{Steinhauer:2003aa} and \cite{Steinhauer:2009aa} indicate that there could be one wave packet satisfying all desiderata.

For WH flows, because of the time reversal symmetry, the situation is the opposite in that there is no difficulty to prepare an initial wave-packet containing only positive frequencies, whereas the final positive and negative frequency packets might be difficult to distinguish as they will follow similar trajectories~\cite{Rousseau:2008aa}.

\section{White holes \label{appWH}}

\begin{figure}
\includegraphics{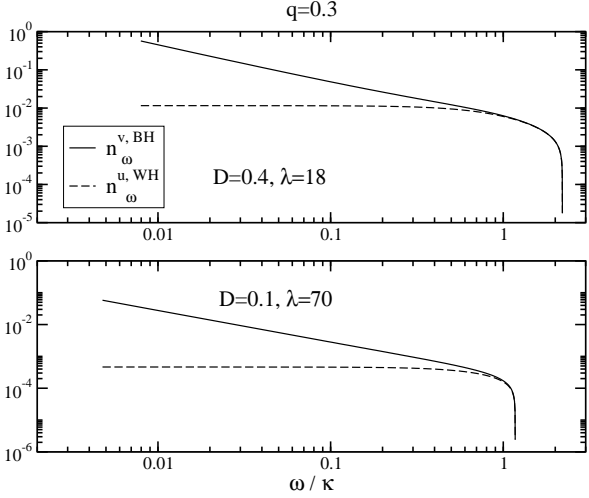}
\caption{Occupation number of the $v$ quanta for a BH ($\bar n^{v}_{\om}$, solid line) and of the $u$ quanta for the corresponding WH under the transformation~\eqref{transfoWHBH} ($\bar n^{\rm WH,u}_{\om}$, dashed line), as functions of $\om/\kappa$ for $q=0.3$ and $D=0.4$ (upper plot) and $D=0.1$ (lower plot).\label{fig::WHB}}
\end{figure}

\begin{figure}
\includegraphics{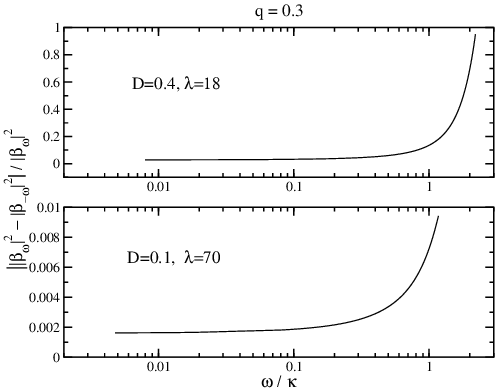}
\caption{Relative difference $\left| |\beta_\om|^2 - |\beta_{-\om}|^2 \right| / |\beta_{\om}|^2$ as a function of $\om$. Same parameters as in \figr{fig::WHB}.\label{fig::WHbeta}}
\end{figure}

So far we only considered condensate flows that possess a sonic horizon which is analogous to that of 
a black hole. 
However sonic horizons which act as a white hole are closely related to the black hole ones. Indeed, 
the transformation
\begin{empheq}{align}
v(x) &\to -v(x), \nonumber\\
 c(x) &\to c(x) , \label{transfoWHBH}
\end{empheq}
transforms the profile studied in the text, 
\eqr{vdparam}, into a white hole profile, in which the characteristics 
focus forward in time, see \eqr{nhr2} with $\kappa < 0$. 
(This equation now applies to the 
left-moving modes, so that the quanta mainly produced are left-moving ones.)

The equivalent of~\eqr{centraleq2td} in this WH profile reads:
\be
\begin{split}
\Big\{\left[ \hbar \left(\partial_t - v \partial_x \right) - i T_\rho \right] \frac{1}{c^{2}} \left[ \hbar \left(\partial_t - v \partial_x \right) + i T_\rho \right] \\
+\, 2m T_\rho \Big\} \phi_j 
 = 0.
\end{split}
\label{centraleq2tdWH}
\ee
Under a time-reversal $t\to -t$, \eqr{centraleq2tdWH} becomes the complex conjugate of \eqr{centraleq2td}. This proves that there is a one-to-one mapping given by:
\be
{\cal T}: \phi_j^{\rm BH}(t,x) \mapsto \phi_j^{\rm WH}(t,x) = \left[\phi_j^{\rm BH}(-t,x)\right]^*.\label{transfoModesBHWH}
\ee
The same relation holds between $\varphi_j^{\rm BH}$ et $\varphi_j^{\rm WH}$. 
The doublets $W_j^{\rm WH}$ and $W_j^{\rm BH}$ related by \eqr{transfoModesBHWH} 
thus have the same norm for the scalar product \eqref{BECnorm}.

With these remarks, it is manifest that in a stationary WH 
flow, when working at fixed frequency $\om$, the modes:
\begin{align}
\phi_{\om}^{v, in, \rm WH}(t,x) &= e^{-i\om t}\left[\phi_\om^{u,out,\rm BH}(x)\right]^*,\nonumber \\
\phi_{-\om}^{v, in, \rm WH}(t,x) &= e^{+i\om t}
\left[\phi_{-\om}^{u,out,\rm BH}(x)\right]^*,\nonumber \\
\phi_{\om}^{u, in, \rm WH}(t,x) &= e^{-i\om t}\left[\phi_\om^{v,out,\rm BH}(x)\right]^*,
\end{align}
(and similarly for $\varphi^{i,in,\rm WH}_\om$) form a complete in basis. 
Notice that the in/out character and the $u$/$v$ 
character of the modes is interchanged between the BH and the WH. 
Indeed, because of the time-reversal symmetry, the phase and group velocities of the modes change sign. 
The Bogoliubov transformation relating the in to the out WH modes 
is thus the inverse of \eqref{Bog}. This implies that  
when dealing with a WH flow in the initial vacuum, instead of \eqr{threeoccn}, 
the mean occupation numbers of emitted quanta are
\begin{align}
\bar n^{\rm WH}_\om &= |\beta_{-\om}|^2, \nonumber \\
\bar n^{u,\rm WH}_{\om} &= |B_\om|^2,\nonumber \\
\bar n^{\rm WH}_{-\om} &= \bar n^{\rm WH}_\om + \bar n^{\rm WH,u}_{\om} = \bar n^{\rm BH}_{-\om}. \label{eqWHBb}
\end{align}
The third line shows that the total number of produced pairs 
is equal to that for the corresponding BH. 
However since $|\beta_\om|^2\neq |\beta_{-\om}|^2$, 
the repartition of the quanta with positive $\om$ into left and right movers differs.

Before considering the implication of these relations, 
it is worth noting that only one quantity governs all differences between the occupation numbers and the elastic scattering in the WH and BH cases. Indeed, 
\eqr{eqWHBb} (or the third equation in \eqref{threeoccn}) shows that
\be
|\beta_\om|^2 + |\tilde B_\om|^2 = |\beta_{-\om}|^2 + |B_\om|^2,
\ee
while the conservation of the norm of $\phi^{u,\rm out, BH}_\om$ and of $\phi^{u,\rm in, BH}_\om$ gives:
\be
|\alpha_\om|^2 + |\tilde A_\om|^2 - |\beta_{-\om}|^2 = |\alpha_\om|^2 + |A_\om|^2 - |\beta_\om|^2 (=1).
\ee
Thus,
\be
|\beta_\om|^2 - |\beta_{-\om}|^2 = |B_\om|^2 - |\tilde B_\om|^2 = |A_\om|^2 - |\tilde A_\om|^2.
\ee
The relative difference between the occupation numbers in the BH and WH cases is small whenever the $u$-$v$ mixing coefficients are small, see for instance Sec.~V C 3 in~\cite{Macher:2009tw}. In the case of a BEC, however, the $u$-$v$ mixing coefficients are not necessarily small and can even grow as $1/\om$ at low frequencies, see Sec.~\ref{numresnv}. 

To further analyze the difference between WH and BH fluxes, in \figr{fig::WHB}, 
we have represented $|B_\om|^2$ and $|\tilde B_\om|^2$ for $q=0.3$ and two values of $(D,\lambda)$. Contrary to $|\tilde B_\om|^2$ that grows as $1/\om$, $|B_\om|^2$ is nearly constant at low frequencies, so that their difference grows like $1/\om$. The consequence of this is that the relative difference $(|\beta_\om|^2 - |\beta_{-\om}|^2) / |\beta_{\om}|^2$ tends to a constant at low frequencies, as is verified in \figr{fig::WHbeta}, where this difference is shown for the same parameters as the previous figure. This constant value can be significant. For $D=0.4$ it is slightly greater than $1\%$; for $D=0.1$, it is very small, of the order of $0.1\%$. For both values of $D$, the relative difference becomes important near $\om_{\rm max}$, but both $|\beta_\om|^2$ and $|\beta_{-\om}|^2$ vanish when reaching this frequency.

From \eqr{eqWHBb} several important lessons can be drawn.
Firstly, we established that, when starting with the vacuum state, 
the fluxes of phonons emitted by a WH flow are directly related to those of the corresponding BH flow. 

Secondly, from this correspondance,  the WH flows appear 
to be as stable as the corresponding BH flows. 
It should be stressed however that this conclusion is reached when using real frequencies $\om$, and
not taking into account the modes that are not asymptotically bounded.  
Whether this implies that WH flows are stable is a moot point which deserves further study,
see \cite{Barcelo:2006yi}.

Thirdly, when expressed in the asymptotic regions in terms of the comoving frequency $\Omega^a(\om)$  
(or the wave number $k^a(\om)$)
the spectral properties of the WH fluxes are radically different from those of the corresponding BH. 
This is because, when considering a WH flow, 
the relationships between the comoving $\Omega^a$ and the conserved frequency $\om$ are time 
reversed with respect to those for a BH. Hence, unlike what was found for a BH flow, \eqr{Omin} 
now determines the final values of the 
$\Omega^a$ which are thus blueshifted. 
In the ideal case of an initial vacuum, this would be a great advantage since the detection of higher frequency phonons is easier. 
When taking into account the nonvanishing temperature of the condensate, 
the initial occupation numbers are now given by $n^{a, \rm in, WH}(\om) = 
\left(e^{\Om^{a, \rm out, BH}(\om)/T_{\rm in}} - 1\right)^{-1}$, hence they are much larger than those in the corresponding BH case since $\Om^{a,\rm out, BH}\simeq \om$. It implies that the initial occupation numbers in WH flows will always be larger than the number of quanta spontaneously created. 
From this, two conclusions can be drawn. First, the detection of the analog Hawking radiation, based on the measurement of final occupation numbers, will be more difficult than in the corresponding BH. Second, on the contrary, the observation of the long-distance correlations of Eqs. \eqref{ldcorr1} and \eqref{ldcorr2} will be greatly facilitated as the larger 
initial occupation numbers increase their amplitude.

Fourthly, WH might also be more appropriate than BH to manipulate coherent states
because at early time, as explained in App.~\ref{coher}, positive and negative $\om$ wave packets follow very different trajectories,
whereas for BH they essentially follow the same trajectory.

\bibliography{../biblio/biblio}
\end{document}